\documentstyle[12pt]{ioplppt}          % use this for preprint style

\def\Lie{\hbox{\it \char'44}\!}                   % Lie derivative

\begin{document}

\jl{6}

\title{Quasilocal quantities for GR and other gravity theories}%
[Quasilocal quantities for gravity theories]

\author{Chiang-Mei Chen\footnote{present address:
Department of Theoretical Physics, Moscow State University, 119899,
Moscow, Russia; email: chen@grg1.phys.msu.su}
and James M. Nester\footnote{email: nester@joule.phy.ncu.edu.tw}}

\address{Department of Physics,
National Central University, Chungli, Taiwan 32054, R.O.C.}

\begin{abstract}
From a covariant Hamiltonian formulation, by using symplectic ideas,
we obtain certain covariant boundary expressions for the quasilocal
quantities of general relativity and other geometric gravity theories.
The contribution from each of the independent dynamic geometric
variables (the frame, metric or connection) has two possible covariant
forms associated with the selected type of boundary condition.  The
quasilocal expressions also depend on a reference value for each
dynamic variable and a displacement vector field.  Integrating over a
closed two surface with suitable choices for the vector field gives
the quasilocal energy, momentum and angular momentum.  For the special
cases of Einstein's theory and the Poincar\'e Gauge theory our
expressions are similar to some previously known expressions and give
good values for the total ADM and Bondi quantities.  We apply our
formalism to black hole thermodynamics obtaining the first law and an
associated entropy expression for these general gravity theories.  For
Einstein's theory our quasilocal expressions are evaluated on static
spherically symmetric solutions and compared with the findings of some
other researchers.  The choices needed for the formalism to associate
a quasilocal expression with the boundary of a region are discussed.

\end{abstract}

\noindent


\maketitle

%%%%%%%%%%%%%%%%%%%%%%%%%%%%%%%%%%%%%%%%%%%%%%%%%%%%%%%%%%%%%%%%%%%%%%
% Introduction                                                       %
%%%%%%%%%%%%%%%%%%%%%%%%%%%%%%%%%%%%%%%%%%%%%%%%%%%%%%%%%%%%%%%%%%%%%%
\section{Introduction}

The fundamental quantities energy-momentum, and angular momentum of
the gravitational field are elusive.  Globally, for spacetimes which
are asymptotically flat (or even anti-DeSitter), there are well
defined values for the {\it total} energy, etc., given by the Bondi
and ADM expressions integrated over spheres at null infinity and
spatial infinity, values which are directly related to quantities that
can be physically measured by distant observers.  However, unlike all
other fields, for the gravitational field these quantities have no
well defined {\it local density} since, according to the {\it
equivalence principle}, an observer cannot detect any features of the
gravitational field at a point.  It has even been argued that the
proper energy-momentum of the gravitational field is only total; that
it cannot (or should not) be localized (see, e.g., \cite{MiThWh73} p
467).

Localization is certainly possible if it is simply understood to mean
``find some way of dividing up the total''.  In particular each of the
many proposed {\it pseudotensors} provides for such a localization.
But pseudotensors depend on the reference system, so they actually
provide for many different localizations each of which includes some
rather arbitrary unphysical content; thus one can arrange, at any
selected point, for almost any value:  positive, zero or even
negative.  Certain positive energy proofs (e.g., \cite{Wi81,
JeKi87, Ne89}) may be a better alternative:  each gives
localizations which are {\it positive}.  But actually these ``positive
localizations'' are not truly ``local'', for they divide up the total
in a way which depends on the configuration nonlocally, since they
each depend on the solution of an elliptic equation which in turn
depends on the field values everywhere.

Appreciating the fundamental nonlocality of the gravitational
interaction yet believing in the basically local nature of physical
interactions led to the idea of {\it quasilocal quantities}:
quantities that take on values associated with a compact orientable
spatial 2-surface.  Expressions for quasilocal quantities in the
context of Einstein's general relativity theory have been proposed
from many perspectives including null rays \cite{Haw68}, twistors
\cite{Pe82}, a fixed background \cite{KaOr90}, symplectic reduction
\cite{JeKi90}, spinors \cite{DoMa91,Be92b,NeTu95,Rob96}, a 2+2
``Hamiltonian'' \cite{HayS94}, Hamilton-Jacobi \cite{BrYo93a} and
Ashtekar variables \cite{Lau93}.  Not surprisingly the various
definitions generally give different results \cite{Be92a,Be92b}.

Lists of criteria to be satisfied by a quasilocal energy have been
devised, see, e.g., \cite{ChYa88,Be92b}.  Usually it is
required that it should vanish for flat spacetime, give reasonable
values for weak fields and spherically symmetric solutions, and should
approach the ADM and Bondi values in the appropriate limits.  Opinions
differ concerning whether quasilocal energy must be non-negative, but
in any case such criteria are not sufficiently restrictive---in fact
there remain an infinite number of possibilities \cite{Be92b}.

Here we offer a comprehensive presentation of some new ideas regarding
quasilocal energy, many of which were first developed in \cite{Ch94}
and briefly reported in \cite{ChNeTu95}.  In our approach we begin
with the idea that energy is naturally associated with time
translation and is thus given by the value of the time translation
generator:  the Hamiltonian.  Hence good expressions for quasilocal
quantities (energy, etc.) should be based not only on the variational
principle and Noether's theorem but especially on the canonical
Hamiltonian procedure, a viewpoint shared by some other investigators,
e.g., \cite{JeKi90,BrYo93a}.

The fundamental feature that distinguishes our Hamiltonian based
expressions for quasilocal quantities (aside from the fact that we
treat rather general geometric gravity theories) is that they are
4-dimensionally covariant.  We believe this is an essential
quality for a physically meaningful (observer independent) definition
of quasilocal quantities appropriate to a covariant gravity theory.
Technically, our {\it covariant} Hamiltonian formulation \cite{Ne91a}
is obtained by using differential form techniques \cite{Ne84}, in lieu
of the usual spacetime splitting with its associated loss of manifest
4-covariance.

For dynamic geometry theories the Hamiltonian has a special form
connected with its role as the generator of displacements along a
timelike vector field $N$.  Noether's theorem applied to local
translations along a vector field reveals that the Hamiltonian 3-form
(density) has the special form ${\cal H}(N) \equiv $ (terms
proportional to field equations) $+ d {\cal B}(N)$.  Consequently, the
Hamiltonian $H(N) = \int_\Sigma {\cal H}(N)$, which displaces a finite
spacelike region $\Sigma$ along $N$, has a value (on a solution to the
field equations) given just by $\oint_{\partial\Sigma} {\cal B}(N)$,
the integral of the 2-form ${\cal B}(N)$ over the boundary of the
spatial region.  Thus this boundary integral will determine the value
of the quasilocal quantities.  Although the Hamiltonian field
equations themselves do not depend on the boundary term, the
expression for ${\cal B}(N)$ (unlike the case for other Noether
conserved currents) is nevertheless restricted by the Hamiltonian
variational principle.

To fix the Hamiltonian boundary term we consider the variation of the
Hamiltonian density.  In general we get an expression of the form
$\delta {\cal H}(N) = $ (field equation terms) $ + d {\cal C}(N)$.
The total differential gives rise to a boundary integral term in the
variation of the Hamiltonian.  As Regge and Teitelboim have nicely
explained \cite{ReTe74}, it is necessary that the boundary term in the
variation of the Hamiltonian vanish (only then are the functional
derivatives well defined).  This is not a problem for finite regions
--- if we fix the appropriate quantities on the boundary.  However,
when we consider the limit $r\to\infty$,the boundary term $\oint {\cal
C}(N)$ for gravity theories does not vanish asymptotically in general.
(In particular this is in fact the case for Einstein's theory.)\ \ To
compensate for this the ${\cal B}(N)$ term in the Hamiltonian needs to
be adjusted.  In this way the form of ${\cal B}(N)$ at infinity is
constrained and the value of the total conserved quantities fixed.

This argument was applied to the Poincar\'e Gauge theory (PGT) and
even more general gravitational theories for both asymptotically flat
spaces and asymptotically constant curvature spaces by one of us
\cite{Ne91a, Ne93} with important improvements by Hecht \cite{He93}.
Expressions for the Hamiltonian boundary integrand which give the
correct total conserved energy, momentum and angular momentum at
spatial infinity \cite{HeNe93} and at future null infinity
\cite{HeNe96} were obtained.  These same boundary expressions can also
be applied to a finite region to give quasilocal values.  But the very
form of Hecht's improvement helped to show the way to other equally
valid expressions.

A comprehensive and systematic investigation of the role of total
differential terms (i.e., boundary terms) in both the Lagrangian and
Hamiltonian variational principle is really needed.  Years ago
Kijowski emphasized the importance of symplectic methods \cite{Ki83}
in such investigations.  More recently there has been increasing
recognition of the importance of symplectic methods and of boundary
terms in variational principles; we have benefited from the progress
made by several workers \cite{Ki84, Yo86,  LeWa90, JeKi92, SuWa92, Wa90,
BrYo93a}.

A nice mathematical theory has been developed, which purports to
incorporate the fundamental principle of physical interactions, and is
especially well adapted to our problem.  This is the theory of {\it
symplectic relations} first proposed by Tulczyjew \cite{Tul74} and
developed with Kijowski \cite{KiTu79}.  (Later Kijowski and coworkers
applied it to gravity theories, e.g.  \cite{JeKi90, Kij96}.  More
recently Wald \cite{Wa93b} has independently obtained some nice
symplectic results.)\ \ In this treatment the (normally neglected)
boundary term in the variational principle reflects the {\it
symplectic structure} which, in turn reveals what physical variables
are the ``control'' variables (held fixed) and which are the
``response'' variables (determined by the physical system).

We shall use this theory to better understand both the general dynamic
geometry Lagrangian variational principle and its associated boundary
terms and the covariant Hamiltonian formulation.  Hence the symplectic
formulation is a key principle underlying our work.  With its aid we
will recognize the symplectic structure of the boundary term ${\cal
C}(N)$ in the variation of the Hamiltonian and be guided to select
those Hamiltonian boundary terms ${\cal B}(N)$ which give rise to a
``covariant'' symplectic structure for ${\cal C}(N)$.

The application of these ideas, the covariant Hamiltonian procedure
along with a covariant symplectic boundary variation, determines the
expressions for the quasilocal quantities of a gravitational system.
In this way we obtain a Hamiltonian 3-form ${\cal H}(N)$, which
generates the evolution along the vector field $N$, and which
includes, for each independent variable (e.g., the connection, the
coframe, the metric), one of two possible covariant boundary terms
depending whether the field or its conjugate momenta is held fixed
(controlled) on the boundary.  Then the variation of the Hamiltonian,
in addition to the field equations, includes a boundary term with a
covariant {\it symplectic} structure which reflects the choice of
control mode.  For each geometric field (metric, frame, connection) we
can control either the field value or the momentum (essentially,
Dirichlet or Neumann boundary conditions).  Each choice gives rise to
a different quasilocal value.  Thus, as in thermodynamics, there are
several different kinds of ``energy'', each corresponds to the work
done in a different (ideal) physical process.

The Hamiltonian boundary terms are the expressions for the quasilocal
quantities.  They determine the value of the Hamiltonian.  The
physical meaning depends on the displacement:  energy for a time
translation, momentum for a space translation, and angular momentum
for a rotation.  (Although our Hamiltonian formalism apparently
presumes a timelike displacement, this is really no limitation, for
the Hamiltonian is linear in the displacement vector field.  Hence
considering the difference between two suitably chosen timelike
displacements gives the meaning of the Hamiltonian for spacelike
displacements such as translations and rotations.)

A noteworthy feature of our expressions is that they have an explicit
dependence on a selected {\em reference configuration}.  The reference
configuration has a simple meaning:  when the dynamic fields have the
reference configuration values on the boundary then all quasilocal
quantities vanish.  In particular, the reference configuration
determines the ``zero'' of energy.

We present the formalism for rather general geometric gravity
theories.  Specifically gravity theories which can have an independent
metric and connection.  The connection need not be metric compatible
nor symmetric.  The field equations are presumed to follow from a
Lagrangian which may depend in any way on the metric, the curvature,
the torsion and non-metricity tensors (but {\it not} on their derivatives).
This general geometric approach has proved to be a good guide for
specific and more specialized theories.  Our expressions are easily
restricted to special cases such as Riemann-Cartan, Riemannian or
teleparallel geometries.  For Einstein's theory and the Poincar\'e
gauge theory (for asymptotically flat {\it or} asymptotically constant
curvature solutions) our expressions are related to previously known
expressions (e.g., \cite{ReTe74, BeMu87, HaSh85, Kaw88, BlVa88}) and
give good values for the total conserved quantities at spatial
\cite{HeNe93} {\it and} future null infinity \cite{HeNe96}.

The outline of this work is:  in section 2, we give the covariant
Hamiltonian analysis of a general geometric gravitational theory in
the language of differential forms.  In section 3 our requirements and
expressions, based on the covariant-symplectic Hamiltonian, for
quasilocal quantities are presented.  We apply the ideas to black hole
thermodynamics obtaining the generalized first law and an expression
for entropy for these general theories in section 4.  The
specialization of our expressions to certain gravity theories is
discussed in section 5.  We restrict our relations to general
relativity, apply them to spherically symmetric solutions and compare
them with some other authors' results in section 6.  Finally our
concluding discussion considers the various choices in selecting a
Hamiltonian boundary term-quasilocal expression, noting in particular
how this scheme provides for an orderly system relating the choices to
physically meaningful properties on the boundary of a spatial region.

%%%%%%%%%%%%%%%%%%%%%%%%%%%%%%%%%%%%%%%%%%%%%%%%%%%%%%%%%%%%%%%%%%%%%%
% Covariant Hamiltonian formalism                                    %
%%%%%%%%%%%%%%%%%%%%%%%%%%%%%%%%%%%%%%%%%%%%%%%%%%%%%%%%%%%%%%%%%%%%%%
\section{Covariant Hamiltonian formalism}

In this section we briefly recount the relevant features of the
covariant Hamiltonian formalism \cite{Ne91a, Ne93}.  We consider quite
general dynamic geometry theories of the metric-affine type (see,
e.g., \cite{HeMcMiNe93}) as well as some important special cases.
The possible geometric potentials are the {\it metric coefficients}
$g_{\mu\nu}$ the {\it coframe} one-form $\vartheta^\alpha$
and the {\it connection} one-form $\omega^\alpha{}_\beta$.
The corresponding field strengths are the {\it non-metricity} one-form
\begin{equation}
Dg_{\mu\nu}:=dg_{\mu\nu}-\omega^{\gamma}{}_{\mu}g_{\gamma\nu}
                        -\omega^{\gamma}{}_{\nu}g_{\mu\gamma},
\end{equation}
the {\it torsion} 2-form
\begin{equation}
\Theta^\alpha := D \vartheta^\alpha := d \vartheta^\alpha +
\omega^\alpha{}_\beta \wedge \vartheta^\beta,
\end{equation}
and the {\it curvature} 2-form
\begin{equation}
\Omega^\alpha{}_\beta := d \omega^\alpha{}_\beta
+ \omega^\alpha{}_\gamma \wedge \omega^\gamma{}_\beta.
\end{equation}

%=====================================================================
%\subsection{Lagrangian formulation}

The usual approach is that (second order) dynamical equations are
presumed to follow from a variational principle.  We presume that the
Lagrangian
density is a scalar valued 4-form depending only on the gauge
potentials and their first differentials:
${\cal L}^{(2)} = {\cal L}^{(2)} (g, \vartheta, \omega; d g, d
\vartheta, d \omega)$.
(Note that we have, in a very natural way, excluded higher derivatives
of the fields. However, there is
really no obstacle in principle to extending our formalism to include
them, say via the expense of introducing extra fields and Lagrange
multipliers.)\ \ In view of the invariance
under local Lorentz transformations the more covariant form:
 ${\cal L}^{(2)} (g, \vartheta; D g,\Theta, \Omega)$ has advantages.
Instead of this usual 2nd order type Lagrangian 4-form we
use a ``first order'' Lagrangian 4-form:
\begin{equation}
{\cal L}
 :=Dg_{\mu\nu} \!\wedge\! \pi^{\mu\nu}
  + \Theta^\alpha \!\wedge\! \tau_\alpha
  + \Omega^\alpha{}_\beta \!\wedge\! \rho_\alpha{}^\beta
  - \Lambda( g, \vartheta; \pi, \tau, \rho).
\label{L1}
\end{equation}
Independent variation with respect to the potentials
$(g,\vartheta,\omega)$ and their conjugate momenta $(\pi,\tau,\rho)$
yields
\begin{eqnarray}
\fl \qquad\delta {\cal L} &=&
 \delta g_{\mu\nu} \, {\delta {\cal L} \over \delta g_{\mu\nu}}
 + \delta \vartheta^\alpha \wedge
   {\delta {\cal L} \over \delta \vartheta^\alpha}
 + \delta \omega^\alpha{}_\beta \wedge
   {\delta {\cal L} \over \delta \omega^\alpha{}_\beta}
+{\delta {\cal L} \over \delta \pi^{\mu\nu}} \wedge \delta \pi^{\mu\nu}
+{\delta {\cal L} \over \delta \tau_\alpha} \wedge \delta \tau_\alpha
\nonumber\\
\fl &+&{\delta {\cal L} \over \delta \rho_\alpha{}^\beta} \wedge
   \delta \rho_\alpha{}^\beta
 + d ( \delta g_{\mu\nu} \pi^{\mu\nu}
     + \delta \vartheta^\alpha \wedge \tau_\alpha
     + \delta \omega^\alpha{}_\beta \wedge \rho_\alpha{}^\beta),
\label{deltaL1}
\end{eqnarray}
which implicitly defines the {\em first order\/} equations
(their explicit form will not be needed).

%=====================================================================
%\subsection{Hamiltonian formulation}

The Hamiltonian formulation displays the ``time'' evolution of the
physical system.  To proceed in a covariant way we consider instants
of ``time'' to be given by spacelike hypersurfaces determined by a
function $t$.  A ``time evolution'' vector field $N$ displaces one
constant $t$ space-like hypersurface $\Sigma_t$ to another;
consequently $i_N dt=1$.  Time evolution is then given by the Lie
derivative where $\Lie_N=i_Nd + di_N$ on the components of form
fields.  We regard $t$ and $N$ as being fixed when we vary our
physical variables.  One way to obtain a set of Hamiltonian equations
is the ``spatial restriction'' (pullback) and ``time projection''
(evaluation on the vector field $N$) of the first order equations.
This would give the initial value constraint and dynamical evolution
equations respectively.  The same equations can also be obtained from
a Hamiltonian.  To obtain the Hamiltonian in a covariant form we
decompose the Lagrangian as follows:
\begin{eqnarray}
{\cal L} &\equiv& dt \wedge i_N {\cal L} \nonumber \\
&=& dt \wedge ( \Lie_N g_{\mu\nu} \pi^{\mu\nu}
  + \Lie_N \vartheta^\alpha \wedge \tau_\alpha
  + \Lie_N \omega^\alpha{}_\beta \wedge \rho_\alpha{}^\beta
  - {\cal H}(N) ).
\label{pq-H}
\end{eqnarray}
In this way we find the {\it Hamiltonian} 3-form
${\cal H}(N) = N^\mu{\cal H}_\mu + d {\cal B}(N)$, where
\begin{eqnarray}\fl\qquad
 N^\mu{\cal H}_\mu &:=& i_N \Lambda + Dg_{\mu\nu} \wedge i_N
\pi^{\mu\nu}   - \Theta^\alpha \wedge i_N \tau_\alpha
  - \Omega^\alpha{}_\beta \wedge i_N \rho_\alpha{}^\beta
  - i_N \vartheta^\alpha \wedge D \tau_\alpha
  \nonumber \\
\fl &-& i_N \omega^\alpha{}_\beta ( D \rho_\alpha{}^\beta
  - g_{\alpha\nu} \pi^{\beta\nu} - g_{\mu\alpha} \pi^{\mu\beta}
  + \vartheta^\beta \wedge \tau_\alpha ), \\
\fl \qquad{\cal B}(N) &:=& i_N \vartheta^\alpha \tau_\alpha
  + i_N \omega^\alpha{}_\beta \, \rho_\alpha{}^\beta.
\label{B}
\end{eqnarray}
The Hamiltonian 3-form ${\cal H}(N)$ is the same quantity which arises
when one considers Noether's second theorem for a local translation
along a vector field $N$ applied to the first order Lagrangian
(\ref{L1}), as we now briefly explain.

Geometric theories are invariant under local diffeomorphisms.
For a first order Lagrangian of the form ${\cal L}=d\varphi\wedge
p-\Lambda$ the general variational formula
\begin{equation}
\delta {\cal L}=d(\delta\varphi\wedge p) +
 \delta\varphi\wedge {\delta {\cal L}\over \delta\varphi}+
 {\delta {\cal L}\over \delta p}\wedge \delta p,
\end{equation}
must be identically satisfied if the variations are
generated by an infinitesimal diffeomorphism. Hence
\begin{equation}
\Lie_N{\cal L}=d i_N {\cal L} \equiv d(\Lie_N\varphi\wedge p) +
 \Lie_N\varphi\wedge {\delta {\cal L}\over \delta\varphi}+
 {\delta {\cal L}\over \delta p}\wedge \Lie_N p.
\end{equation}
Consequently the ``Hamiltonian 3-form''
\begin{equation}
{\cal H}(N):=\Lie_N\varphi\wedge p-i_N{\cal L},
\label{Ham}
\end{equation}
satisfies a differential identity
 \begin{equation}
d{\cal H}(N)
\equiv{\hbox{(terms proportional to field equations)}}.
\label{dH=fe}
\label{dH}
\end{equation}
In addition to this Noether differential identity, because displacement
along $N$ is a local symmetry, we also have an algebraic identity.
Using ${\cal L}=d\varphi\wedge p-\Lambda$  in (\ref{Ham})
we rearrange the Hamiltonian into the form
${\cal H}(N)=N^\mu{\cal H}_\mu+d{\cal B}(N)$ and then substitute
$d{\cal H}(N)=d(N^\mu{\cal H}_\mu)=dN^\mu\wedge{\cal H}_\mu+N^\mu
d{\cal H}_\mu$
into the differential identity (\ref{dH=fe}).  The coefficient of
$dN^\mu$ gives an algebraic identity,
 \begin{equation}
{\cal H}_\mu \equiv{\hbox{(terms proportional to field equations)}}.
\label{Hmu}
\end{equation}
From (\ref{dH}) we see that ``on shell'' (i.e., when the field
equations are satisfied) the Hamiltonian 3-form is a ``conserved
current'' and the value of the Hamiltonian $H(N)$, the integral of the
${\cal H}(N)$ over a space-like hypersurface $\Sigma$, is a conserved
quantity.  From the algebraic identity (\ref{Hmu}), ${\cal H}_\mu$
vanishes when evaluated for solutions of the field equations.  Thus,
``on shell'', the value of Hamiltonian $H(N)$ depends only on the term
$d {\cal B}(N)$.  By the generalized Stokes theorem, the value of the
Hamiltonian is just the integral of the 2-form ${\cal B}(N)$ over the
2-surface boundary $\partial \Sigma$ of the 3-hypersurface $\Sigma$.
This value should determine the {\it energy} and the other quasilocal
quantities.  However, just as with other Noether currents,
${\cal H}(N)$ is not unique; we can add to it a total differential
without affecting the conservation relation (\ref{dH}), this amounts
to modifying ${\cal B}(N)$ (\ref{B}).  Such a modification does not
affect the Hamiltonian equations of motion.  The Hamiltonian field
equations are obtained by varying the action (\ref{pq-H}).  The
variation of the above Hamiltonian has the form
\begin{equation}
\delta {\cal H}(N) = \hbox{(field equation terms)} + d {\cal C}(N),
\end{equation}
where
\begin{equation}
{\cal C}(N) := i_N (\delta g_{\mu\nu} \pi^{\mu\nu}
  + \delta \vartheta^\alpha \wedge \tau_\alpha
  + \delta \omega^\alpha{}_\beta \wedge \rho_\alpha{}^\beta).
\label{C(N)}
\end{equation}
A modification of ${\cal B}(N)$ (\ref{B}) affects only the total
derivative term $d{\cal C}(N)$ which, upon integration over $\Sigma$
becomes a boundary term.

%%%%%%%%%%%%%%%%%%%%%%%%%%%%%%%%%%%%%%%%%%%%%%%%%%%%%%%%%%%%%%%%%%%%%%
% Expressions for Quasilocal Quantities                              %
%%%%%%%%%%%%%%%%%%%%%%%%%%%%%%%%%%%%%%%%%%%%%%%%%%%%%%%%%%%%%%%%%%%%%%
\section{Expressions for quasilocal quantities}

Our purpose is to find quasilocal quantities for gravity theories.
Hence we want to determine the expression for the term ${\cal B}(N)$
in the Hamiltonian.  From the variational principle we know that a
total differential term can be added to the Hamiltonian without
changing the field equations.  But this, of course, will change the
expression of ${\cal B}(N)$, and thereby the value of the quasilocal
quantities.  What is the proper ``physical'' expression for quasilocal
quantities?  We have already noted that the requirement of having the
desired limiting values is far too weak.  We use certain additional
theoretical criteria involving the boundary term in the variation of
the Hamiltonian to severely restrict the form of
a ``good''  expression for the term ${\cal B}(N)$.

%=====================================================================
\subsection{Requirements}

Consider the total differential term ${\cal C}$ in the variation of
the Hamiltonian.  We propose three theoretical requirements on the
term ${\cal C}$ in order to limit the quasilocal boundary expression
${\cal B}$:

\begin{itemize}
\item {\bf [R1] Well-defined Requirement}
  \begin{itemize}
  \item {\sl The term
        ${\cal C}$ must vanish at the boundary in general both for
        finite and infinite regions.}
  \end{itemize}
\item {\bf [R2] Symplectic Structure Requirement}
  \begin{itemize}
  \item {\sl The term ${\cal C}$ must reflect the {\it symplectic}
        structure \cite{KiTu79}, which tells us which physical
        variables are ``control'' variables (i.e., we control them at
        the boundary), and which are ``response'' variables (i.e., are
        determined by physical laws).}
  \end{itemize}

\item {\bf [R3] Covariant Requirement}
  \begin{itemize}
\item {\sl The control and response variables must each
  appear in the form of a covariant combination\/} (as befits a
covariant theory).    \end{itemize}
\end{itemize}
With respect to the Well-defined Requirement R1, the expression of the
term ${\cal C}(N)$ in equation (\ref{C(N)}) is no problem for a finite
boundary.  We need merely fix the appropriate variables to vanish at
the boundary.  For a boundary at infinity, however, one needs to
consider a limit and the natural rate of fall off in the asymptotic
region.  In this case, the term ${\cal C}(N)$ in equation (\ref{C(N)})
is nonvanishing generally.  (Specific examples are the Einstein theory
\cite{ReTe74} and the Poincar\'e gauge theory (PGT) for solutions
which have asymptotically zero or constant curvature.)\ \ In order to
correct ${\cal C}(N)$ to satisfy R1, one {\em must} modify the term
${\cal B}$ in equation (\ref{B}).

%=====================================================================
\subsection{Well-defined version}

Some time ago within the context of metric compatible theories, i.e.,
the PGT, one of us \cite{Ne91a} proposed a modified version of the
boundary term (\ref{B}):
\begin{equation}
{\cal B}(N) = i_N \vartheta^\alpha \Delta \tau_\alpha
+ i_N \omega^\alpha{}_\beta \, \Delta \rho_\alpha{}^\beta
+ \Delta \omega^\alpha{}_\beta \wedge i_N \rho_\alpha{}^\beta.
\label{B0}
\end{equation}
With this adjustment the total differential term ${\cal C}(N)$ in the
variation of ${\cal H}(N)$ takes the form
\begin{equation}
\fl\qquad {\cal C}(N) = \delta i_N \vartheta^\alpha \Delta
\tau_\alpha - \delta \vartheta^\alpha \wedge i_N \tau_\alpha
+ \delta i_N \omega^\alpha{}_\beta \, \Delta
\rho_\alpha{}^\beta + \Delta \omega^\alpha{}_\beta \wedge \delta
i_N \rho_\alpha{}^\beta,
\label{C(N)1}
\end{equation}
where, for any quantity $\alpha$, ${\buildrel \scriptstyle \circ
\over \alpha}$ is the {\it reference configuration} value of $\alpha$
which is fixed under variation, and
$\Delta \alpha := \alpha - {\buildrel \scriptstyle \circ \over
\alpha}$.
This expression should vanish at spatial infinity, in particular for
asymptotically flat or constant curvature PGT solutions.  But for some
theories, e.g., the PGT with a massless torsion field \cite{Ya93}, the
boundary term ${\cal B}(N)$ still needs adjustment.  Hecht recognized
this and proposed an important modification \cite{He93} to the
$i_N\omega$ factor.  (We have already noted that the Hecht expression
has since been successfully tested at spatial {\em and} null infinity
for both asymptotically flat {\em and} asymptotically constant
(negative) curvature solutions \cite{HeNe93, HeNe96}.)

However we can also add a term
 $\Delta \vartheta^\alpha \wedge i_N \tau_\alpha$
 into the original expression (\ref{B0}) to give the
formula more symmetry.  Then
\begin{equation}
{\cal B}'(N) = i_N \vartheta^\alpha \Delta \tau_\alpha
+ \Delta \vartheta^\alpha \wedge i_N \tau_\alpha
+ i_N \omega^\alpha{}_\beta \, \Delta \rho_\alpha{}^\beta
+ \Delta \omega^\alpha{}_\beta \wedge i_N \rho_\alpha{}^\beta,
\end{equation}
and
\begin{equation}
\fl \qquad{\cal C}'(N) = \delta i_N \vartheta^\alpha \Delta
  \tau_\alpha + \Delta \vartheta^\alpha \wedge \delta i_N
  \tau_\alpha + \delta i_N \omega^\alpha{}_\beta \, \Delta
  \rho_\alpha{}^\beta + \Delta \omega^\alpha{}_\beta \wedge \delta
  i_N \rho_\alpha{}^\beta.
\label{C(N)'}
\end{equation}
From these considerations we see that the Well-Defined requirement
R1 cannot determine the quasilocal term ${\cal B}(N)$ uniquely.
Moreover neither ${\cal C}(N)$ nor ${\cal C}'(N)$ satisfy the
Covariant Requirement R3.

%=====================================================================
\subsection{Covariant symplectic structure version}

More recently we have improved these expressions and obtained several
different boundary terms, all of them satisfy the requirements R1, R2
and R3.  They include, for each independent k-form field $\varphi$
(e.g., the connection, the coframe, the metric), one of two possible
covariant boundary terms:
\begin{equation}
\fl\quad {\cal B}_\varphi(N) =
 i_N\varphi \wedge \Delta p -(-1)^k \Delta \varphi \wedge
 i_N {\buildrel\scriptstyle \circ \over p}
 \quad\ \hbox{or} \quad\
{\cal B}_p(N) =
i_N {\buildrel \scriptstyle \circ \over \varphi} \wedge \Delta p
-(-1)^k \Delta \varphi \wedge i_N p,
\label{Bphi}
\label{Bp}
\end{equation}
depending upon whether $\varphi$ or its conjugate momenta $p$
is held fixed (controlled) on the boundary.
Then the variation of the Hamiltonian, in addition to the field
equations, includes a boundary term which is a projection onto the
boundary of a covariant {\it symplectic} structure:
\begin{equation}
d i_N (\delta \varphi \wedge \Delta p) \qquad
\hbox{or} \qquad d i_N (-\Delta \varphi \wedge \delta p),
\label{symp}
\end{equation}
 respectively, which reflects the choice of control mode.

Thus for the geometric fields we take
\begin{eqnarray}
{\cal B}(N) &=&
\left\{ \begin{array}{c}
 -\Delta g_{\mu\nu} \,
i_N{\buildrel\scriptstyle\circ\over\pi}{}^{\mu\nu}\\
 -\Delta g_{\mu\nu} \, i_N \pi^{\mu\nu}
 \end{array} \right\}
+
\left\{ \begin{array}{c}
 i_N \vartheta^\alpha \Delta \tau_\alpha
   + \Delta \vartheta^\alpha \wedge i_N
   {\buildrel \scriptstyle \circ \over \tau}_\alpha \\
 i_N {\buildrel \scriptstyle \circ \over \vartheta}{}^\alpha
   \Delta \tau_\alpha
   + \Delta \vartheta^\alpha \wedge i_N \tau_\alpha
 \end{array} \right\}  \nonumber \\
&+&
\left\{ \begin{array}{c}
 i_N \omega^\alpha{}_\beta \, \Delta \rho_\alpha{}^\beta
   + \Delta \omega^\alpha{}_\beta \wedge i_N
   {\buildrel \scriptstyle \circ \over \rho}{}_\alpha{}^\beta \\
 i_N {\buildrel \scriptstyle \circ \over \omega}{}^\alpha{}_\beta
  \, \Delta \rho_\alpha{}^\beta
   + \Delta \omega^\alpha{}_\beta
   \wedge i_N \rho_\alpha{}^\beta
 \end{array} \right\}\ ,
\label{B(N)}
\end{eqnarray}
where the upper (lower) line in each bracket is to be selected if
the field (momentum) is controlled.
Hence, as in thermodynamics, there are several kinds of ``energy'',
each corresponds to the work done in a different (ideal) physical
process \cite{KiTu79}, \cite{JeKi90}.

For the geometric variables, the total differential
term in $\delta {\cal H}(N)$ is of the boundary projection form $d
i_N {\cal C}$ with $\cal C$ now given by the covariant expression
\begin{equation}
{\cal C}=
 \left\{ \begin{array}{c}
   \delta g_{\mu\nu} \, \Delta \pi^{\mu\nu} \\
   - \Delta g_{\mu\nu} \, \delta \pi^{\mu\nu}
   \end{array} \right\}
+
 \left\{ \begin{array}{c}
   \delta \vartheta^\alpha \wedge \Delta \tau_\alpha \\
   - \Delta \vartheta^\alpha \wedge \delta \tau_\alpha
   \end{array} \right\}
+
 \left\{\begin{array}{c}
\delta \omega^\alpha{}_\beta \wedge \Delta \rho_\alpha{}^\beta \\
  - \Delta \omega^\alpha{}_\beta \wedge \delta \rho_\alpha{}^\beta
   \end{array} \right\} , \label{C}
\end{equation}
where again the upper (lower) line in each bracket corresponds to
controlling the field (momentum).

Of course we cannot expect to get completely covariant
expressions when we have a non-covariant dynamic variable such as a
connection. Note however that $\Delta\omega$, being the difference
between two connections, is tensorial,
so our quasilocal boundary expressions are {\it covariant}---aside
from the manifestly non-covariant explicit connection terms in
${\cal B}$.  These $i_N\omega$ terms include a real covariant physical
effect plus an unphysical (noncovariant) dynamical reference frame
effect.  These two effects can be separated by using the identity
\begin{equation}
(i_N \omega^\alpha{}_\beta) \vartheta^\beta \equiv
 i_N \Theta^\alpha + D N^\alpha
 - \Lie_N \vartheta^\alpha, \label{inomega}
\end{equation}
to replace the $i_N\omega$ factor within the last bracket in
(\ref{B(N)})
with two covariant terms plus a manifestly non-covariant term of the
form $(\Lie_N\vartheta^\alpha)_\beta \Delta\rho_\alpha{}^\beta$.
This $\Lie_N\vartheta \Delta \rho$ piece is really needed in
our general Hamiltonian to generate the dynamics due to the
gauge freedom of the reference frame --- but
it represents an unphysical (observer dependent) contribution to the
quasilocal quantities.  For the purposes of calculating the
physical value of
the quasilocal quantities it should be dropped.  (Note that if $N$ is
a Killing vector then this term can be made to vanish for a suitable
choice of frame.)

%=====================================================================
\subsection{Uniqueness}

An important property of our expressions is {\it uniqueness}.  Of
course, the uniqueness property depends on the proposed requirements.
Under our three requirements (R1, R2 and R3), our
expressions are the only possible ones.

According to our covariant symplectic structure requirement (R2 +
R3), for each field potential $\varphi$ and its momenta $p$, the
boundary term of  the variation of the Hamiltonian
must be one of the following two types
\begin{equation}
d i_N (\delta \varphi \wedge \Delta p) \qquad \hbox{or} \qquad
d i_N (-\Delta \varphi \wedge \delta p).
\end{equation}

The only acceptable modification
to the Hamiltonian which preserves the covariant symplectic structure
requirement is of the form $di_N{\cal F}$, a projection of a
4-covariant form onto the boundary.
The 3-form ${\cal F}$ must depend on the control variables
only algebraically%  ----- Begin footnote -----
\ftnote{1}{Since if it depends on the differential of the control
variable, we must fix the differential of this variable at the
boundary. This is forbidden by the symplectic structure.}
%  ----- End footnote -----
and must be scalar-valued.
Working with our dynamic variable pairs
 $(g_{\mu\nu},\pi^{\mu\nu})$,
$(\theta^\alpha, \tau_\alpha)$ and $(\omega^\alpha{}_\beta,
\rho_\alpha{}^\beta)$,
 the only acceptable combination  is of the form
$d i_N (\Delta \varphi\wedge \Delta p)$.  This just switches
between the two control modes:
\begin{equation}
{\cal B}_p={\cal B}_\varphi-i_N (\Delta \varphi\wedge \Delta p).
\end{equation}
Of course one could add some covariant quantity which doesn't depend
on the dynamic variable.  That would leave the Hamiltonian variation
symplectic structure unchanged but would ``renormalize'' the
quasilocal quantities---albeit in an ``unphysical way''.   We rule
this out by normalizing our quasilocal expressions so that they
vanish if the dynamic variables equal the reference values on the
boundary.

Although we have determined the expressions `uniquely', there still
exist two undetermined things.  One is the displacement vector field
$N$.  For example, if we want to calculate the energy of a physical
system in a finite region, we have not yet specified how to select the
appropriate timelike vector $N$.  The other is the reference
configuration.  We will look closer at these issues later.

%=====================================================================
\subsection{Spacetime version}

Now let's go back to the Lagrangian level using the same
requirements in order to get a completely covariant analysis.
In equation (\ref{deltaL1}), the boundary term satisfies the Covariant
Symplectic Structure Requirement R2 and R3,
but not the Well-defined Requirement R1.

For the purpose of getting boundary terms satisfying our requirements,
we considered ways to modify the Lagrangian by adding an extra total
differential term.  In a first order Lagrangian we found that we could
include a boundary term by letting
\begin{equation}
{\cal L} := d\varphi\wedge p - \Lambda - d {\cal K},
\end{equation}
where ${\cal K}=\Delta \varphi \wedge
 {\buildrel \scriptstyle \circ \over p} $
 or ${\cal K} = \Delta \varphi \wedge p $.
Then
\begin{equation}
\delta {\cal L} \!=\! \hbox{(field equations)}
\!+\! d \biggr(
 \left\{\!\! \begin{array}{c}
   \delta \varphi \wedge \Delta p \\
   -\Delta \varphi \wedge \delta p
   \end{array}\!\! \right\}
\biggr).
\end{equation}
We can relate this to the Hamiltonian analysis for
\begin{equation}
\delta i_N {\cal L} \equiv i_N \delta {\cal L}
 = i_N \hbox{(field equations)} \!+\!
 i_N d \biggr(
 \left\{\!\! \begin{array}{c}
   \delta \varphi \wedge \Delta p \\
   -\Delta \varphi \wedge \delta p
   \end{array}\!\! \right\}  \biggr).
\end{equation}
The latter term can be rearranged into
\begin{equation}
 \Lie_N
\biggr(   \left\{\!\! \begin{array}{c}
   \delta \varphi \wedge \Delta p \\
   -\Delta \varphi \wedge \delta p
   \end{array}\!\! \right\}    \biggr)
 - d i_N
\biggr(   \left\{\!\! \begin{array}{c}
   \delta \varphi \wedge \Delta p \\
   -\Delta \varphi \wedge \delta p
   \end{array}\!\! \right\}  \biggr),
\end{equation}
which consists of a total time derivative (which integrates into a
term at the initial and final times and constitutes a canonical
transformation) and a spatial boundary term which is identical to our
covariant symplectic Hamiltonian boundary variation term.
By the way, it is here that one can see the covariant origin of our
required form for the projection of a covariant quantity (\ref{symp}).

%%%%%%%%%%%%%%%%%%%%%%%%%%%%%%%%%%%%%%%%%%%%%%%%%%%%%%%%%%%%%%%%%%%%%%
%  Black hole thermodynamics                                         %
%%%%%%%%%%%%%%%%%%%%%%%%%%%%%%%%%%%%%%%%%%%%%%%%%%%%%%%%%%%%%%%%%%%%%%
\section{Black hole thermodynamics}

Here we will use arguments similar to those of Brown and York
\cite{BrYo93b, Bro95} and of Wald \cite{Wa93a} (see also
\cite{HaHo96}) along
with our symplectic Hamiltonian expressions to derive the first law of
black hole thermodynamics for these general geometric gravity
theories.

For the theories of gravity that we are considering, the null rays and
causal structure are governed only by the metric and its associated
Riemannian geometry.  Hence the basic geometry of black holes in these
theories should be the same as that of GR.

The quantity
corresponding to the temperature in black hole thermodynamics is the
{\it surface gravity}, $\kappa$, defined at any point of a Killing
horizon by \cite{Wa84}
\begin{equation}
\chi^\alpha \, \nabla_\alpha \, \chi^\beta =: \kappa \, \chi^\beta,
\end{equation}
where $\chi^\alpha$ is the Killing vector field normal to the Killing
horizon and $\nabla$ is the Riemannian covariant derivative.  Hawking
proved that the event horizon of a stationary black hole is a Killing
horizon in general relativity \cite{HaEl73}.  For certain general
gravity theories it has been shown that if a black hole has a
bifurcate Killing horizon (the Killing vector vanishes at this
horizon) then the constancy of the surface gravity holds
\cite{RaWa92}.  For the theories under consideration here we presume
that the black hole has a bifurcate Killing horizon and that the
zeroth law is satisfied:  the surface gravity (which is proportional
to the temperature) is constant over the entire horizon
\cite{JaKaMy93}.

The first law of thermodynamics concerns the conservation of energy.
Because our Hamiltonians satisfy a conservation law, the corresponding
formula for gravitating systems, in particular black holes, can be
obtained and the quantity corresponding to the entropy identified.

The basic idea is to consider the region between the horizon and
infinity.  We wish to vary certain parameters such as the total energy
and angular momentum.  However in the formulations discussed earlier
such quantities were not control parameters.  To obtain the necessary
additional free variations on the boundary we must allow the
displacement vector field to vary.  Then the variation of the
Hamiltonian 3-form, including the variation of the displacement vector
field, is of the form
\begin{eqnarray}
\delta {\cal H}(N)&=& \hbox{(field equation terms)}
+ (\delta N)^\mu{\cal H}_\mu \nonumber\\
& +& d \bigl(i_N (\hbox{symplectic control mode terms})
  + {\cal B}(\delta N) \bigr)\, ,
\end{eqnarray}
with the ${\cal H}_\mu$ term vanishing by the initial value
constraints.  Note that the response variable for the displacement
vector is the ``component'' of the quasilocal quantities.  Here ${\cal
B}$ can be any one of our boundary terms representing any control mode
as long as the ``reference frame energy'' is removed by dropping the
$\Lie_N\vartheta$ contribution via the $i_N\omega$ replacement
(\ref{inomega}).  Following Brown and York \cite{BrYo93b},
 we now introduce the
{\it microcanonical\/} Hamiltonian via a Legendre transformation:
\begin{equation}
{\cal H}_{\mathrm{micro}}(N) = {\cal H}(N) - d {\cal B}(N).
\end{equation}
The variation of the microcanonical action is of the form
\begin{eqnarray} \fl\qquad
\delta (S_{\mathrm{micro}}[N]) &=&\int_{\Sigma}
 \hbox{(field equation terms)}
- (\delta N)^\mu{\cal H}_\mu \nonumber\\
\fl\qquad
&- & \oint_{\partial\Sigma}
 \bigl(i_N (\hbox{symplectic control mode terms})
  + (\delta{\cal B})(N) \bigr)\, .
\end{eqnarray}
The variation of $S_{\mathrm{micro}}$ should vanish on a solution with
the control variables fixed at the boundary.  Suppose the hypersurface
includes a boundary with two components, one at infinity and the other
at the bifurcate Killing horizon, $H$.  Then we get a kind of
energy-momentum conservation law:
\begin{equation}
0=\oint_{\partial\Sigma} (\delta{\cal B})(N)=
\oint_\infty (\delta{\cal B})(N)-
\oint_H (\delta{\cal B})(N)\, .
\end{equation}
Now we shall choose our variations to be such that they perturb the
asymptotic values of the total quantities.

Let $X_t$ and $X_\varphi$ denote the Killing vector fields which
approach the time translation and rotation at the infinity.  Then the
total energy is the value of $H(X_t)=\oint_\infty {\cal B}(X_t)$ and
the total angular momentum is the value of $H(X_\varphi)=\oint_\infty
{\cal B}(X_\varphi)$.  It is worth remarking that this is true for all
the possible control modes, asymptotically they give the same total
results for they differ by terms with more rapid fall off.

We assume as usual that the Killing vector $\chi$ can be
decomposed into
 \begin{equation}
\chi = X_t + \Omega_H X_\varphi,
\end{equation}
and vanishes on the bifurcate Killing horizon, where $\Omega_H$ is the
``angular velocity of the horizon''.

 Now let the displacement
$N$ be the Killing vector $\chi$.  Then
\begin{equation}
\oint_{\infty} (\delta {\cal B})(N) = \delta E + \Omega_H \, \delta J.
\end{equation}
On the other hand, for the surface integration over the
horizon boundary, because the Killing field vanishes on the horizon
the only terms that can contribute involve
 ${\buildrel \scriptstyle \circ \over D}\chi$ or
 $D\chi={\buildrel \scriptstyle \circ \over D}\chi+\Delta\omega\chi$,
both of which which reduce to $\nabla\chi$ on the horizon (since the
affine connections differ from the Levi-Civita connection by tensor
terms multiplied by the vector field which vanishes on $H$).  Hence we
get
 \begin{equation}
\oint_H (\delta{\cal
B})(\chi)=\oint_H \nabla_\beta\chi^\alpha\delta\, \rho_\alpha{}^\beta=
\kappa \oint_H
\epsilon_\beta{}^\alpha \, \delta \rho_\alpha{}^\beta =: \kappa\,\delta
S, \end{equation}
here we have assumed that $\nabla_\alpha \chi_\beta = \kappa
\epsilon_{\alpha\beta}$, where $\epsilon_{\alpha\beta}$ is the
bi-normal to the horizon.  This has the form of the first law and
identifies an expression for the {\it black hole entropy} $S$ for our
general gravity theories to be dependent on the horizon components of
$\rho_\alpha{}^\beta$, which is the field momentum 2-form conjugate to
the curvature.  This result is consistent with the well known Einstein
theory value, where the integral of
$\epsilon^{\alpha\beta}\rho_{\alpha\beta}$ is proportional to the
area.  Remarkably, it is of the same form as an expression found
previously for other general gravity theories which were based on
Riemannian geometry with no torsion or non-metricity but, on the other
hand, were allowed to contained arbitrarily high derivatives of the
curvature \cite{JaKaMy93,Wa93a,Bro95}.  These works used various
techniques ranging from the Noether-Charge analysis of Iyer and Wald
to the microcanonical functional integral method of Brown.  The extent
to which these results can be combined with our work is not yet clear
to us.

%%%%%%%%%%%%%%%%%%%%%%%%%%%%%%%%%%%%%%%%%%%%%%%%%%%%%%%%%%%%%%%%%%%%%%
% Some Theories of Gravity                                           %
%%%%%%%%%%%%%%%%%%%%%%%%%%%%%%%%%%%%%%%%%%%%%%%%%%%%%%%%%%%%%%%%%%%%%%
\section{Some theories of gravity}

Our general formalism readily specializes to
either coordinate
or orthonormal frames, moreover it is suitable for the most general
metric-affine
gravity theories \cite{HeMcMiNe93,GroF97}. (For earlier proposals for
quasilocal quantities for such theories see \S5.7 in
\cite{HeMcMiNe93}.) \ \
Nevertheless, it also readily specializes to less
general geometries like Riemann-Cartan, Riemannian or teleparallel
and to gravity theories formulated in such geometries.

Teleparallel theories (i.e., curvature vanishes), see,
e.g., \cite{HaSh79,MuGrHe98}, are somewhat special; hence we leave
the treatment of their covariant Hamiltonian formalism, and
quasilocal quantities for future work.
For the Poincar\'e gauge theory (PGT) \cite{He80,HaSh80} one
could impose the metric compatible connection condition via a Lagrange
multiplier.  However, a neater method is to use orthonormal frames and
restrict the connection algebraically:
$\omega^{\alpha\beta}=-\omega^{\beta\alpha}$.  Then one need merely
drop all the $g,\pi$ terms from our expressions.  The most general
potential $\Lambda(\vartheta,\tau,\rho)$ contains 3 independent
quadratic terms in $\tau$ and 6 independent quadratic terms in $\rho$:
 \begin{equation} \tau_\alpha =
 -{1\over\chi}* (\sum_{n=1}^3 a_n{}^{(n)}  \Theta_\alpha), \qquad
\rho^{\alpha\beta} = -{a_0 \over 2\chi}\eta^{\alpha\beta}
 -{1\over \kappa}* (\sum_{n=1}^6 b_n {}^{(n)} \Omega^{\alpha\beta})\, ,
\end{equation}
which express the momenta linearly in terms of the algebraically
irreducible parts of the torsion and curvature.  Here we have
introduced the convenient notation
$\eta^{\alpha\dots}:=*(\vartheta^\alpha\wedge\cdots)$, with
$\eta:=*1=
\vartheta^0\wedge\vartheta^1\wedge\vartheta^2\wedge\vartheta^3$.

The Einstein-Cartan theory is a special case of the PGT with
the Lagrangian
\begin{equation}
{\cal L}_{EC} =  {1 \over 2\chi} \Omega^\alpha{}_\beta \wedge
\eta_\alpha{}^\beta + {\Lambda_{cos}\over \chi} \eta\, ,
\label{LEC}
 \end{equation}
(where $\Lambda_{cos}$ is the cosmological constant and $\chi:=8\pi
G/c^4=8
\pi$ in geometric units).  The variables are the orthonormal frame and the
metric compatible (antisymmetric) connection one-form.  The associated
conjugate momenta are
\begin{equation}
\tau_\alpha = 0, \qquad
\rho_\alpha{}^\beta = {1 \over 2\chi} \eta_\alpha{}^\beta\, .
\label{ecmom}
\end{equation}
Properly we should introduce these momenta and regard (\ref{ecmom}) as
constraints which can be enforced with Lagrange multipliers.  However,
because these ``momenta constraints'' are purely algebraic, we can
take a short cut and directly use (\ref{LEC}) as our first order
Lagrangian.  The corresponding quasilocal expressions are
\begin{eqnarray} {\cal B}_\omega  &=&  {1 \over 2 \chi}
 ( \Delta \omega^\alpha{}_\beta \wedge i_N  {\buildrel
 \scriptstyle \circ \over \eta}{}_\alpha{}^\beta +
 i_N \omega^\alpha{}_\beta\, \Delta \eta_\alpha{}^\beta )\, ,
 \label{Bomega} \\
{\cal B}_\vartheta  &=&  {1 \over 2 \chi}
 ( \Delta \omega^\alpha{}_\beta \wedge i_N \eta_\alpha{}^\beta +
i_N {\buildrel \scriptstyle \circ \over
 \omega}{}^\alpha{}_\beta \, \Delta \eta_\alpha{}^\beta ) \, .
\label{Btheta}
\end{eqnarray}

The gravity theory which attracts the most interest is Einstein's
General Relativity.  The Riemannian geometry of GR does not have an
independent connection while our formalism is geared to independent
variations of the frame, metric and connection, Yet there is no
incompatibility, indeed one can proceed in many ways including:
(1) impose the metric compatible and vanishing torsion
conditions via Lagrange multipliers,
(2) use coordinate frames, algebraically impose the
symmetry of the connection coefficients and impose metric
compatibility via a Lagrange multiplier,
(3) use coordinate frames, algebraically impose the
symmetry of the connection coefficients and (for vacuum
and all non-derivative coupled sources) obtain the metric
compatibility condition as a field equation via a Palatini type
variation,
(4) use the orthonormal frame EC theory and
impose the vanishing torsion condition via a Lagrange multiplier,
(5) use the orthonormal frame EC theory and obtain (for vacuum and
all non-derivative coupled sources) the vanishing torsion condition as
a field equation.

The latter method is technically the simplest and we will use it here
to obtain our quasilocal quantities for GR.  All of the other
procedures give similar results.  Similar but not necessarily
identical---because the boundary control differs---for example
 $\Delta\omega^\alpha{}_\beta\wedge\delta\eta_\alpha{}^\beta=
\Delta\omega^\alpha{}_\beta\wedge
\delta\vartheta^\mu\wedge\eta_\alpha{}^\beta{}_\mu$ in terms of
orthonormal (co)frames, but it equals
$\Delta\Gamma^\alpha{}_{\beta\gamma}\,\delta(\sqrt{-g}g^{\beta\sigma})\,
\delta^{\nu\gamma}_{\alpha\sigma}\, d^3x_\nu$ in a coordinate basis.
A comprehensive investigation covering all these options is
underway.

%%%%%%%%%%%%%%%%%%%%%%%%%%%%%%%%%%%%%%%%%%%%%%%%%%%%%%%%%%%%%%%%%%%%%%
% Quasilocal Values in General Relativity                            %
%%%%%%%%%%%%%%%%%%%%%%%%%%%%%%%%%%%%%%%%%%%%%%%%%%%%%%%%%%%%%%%%%%%%%%
\section{Quasilocal values in general relativity}

We now consider our quasilocal expressions for general relativity.
This will help to give a better understanding of their mathematical
nature and physical meaning and will permit a comparison with the
results obtained by other authors.

The Hilbert Lagrangian (\ref{LEC}) of the Einstein(-Cartan) theory
(dropping the cosmological constant and often the constant $2\chi$ for
simplicity) can be spacetime decomposed according to
\begin{eqnarray}
\fl\qquad{\cal L} &=& dt \wedge i_N {\cal L}  \nonumber \\
\fl\qquad&=& dt \wedge [\Lie_N \omega^\alpha{}_\beta \wedge
\eta_\alpha{}^\beta
 + N^\mu \Omega^\alpha{}_\beta \wedge \eta_\alpha{}^\beta{}_\mu
 + i_N \omega^\alpha{}_\beta\, D \eta_\alpha{}^\beta
 - d (i_N \omega^\alpha{}_\beta\, \eta_\alpha{}^\beta)]\, ,
\end{eqnarray}
which identifies the Hamiltonian 3-form as
\begin{equation}
{\cal H} = -N^\mu \Omega^\alpha{}_\beta \wedge
\eta_\alpha{}^\beta{}_\mu
- i_N \omega^\alpha{}_\beta\, D \eta_\alpha{}^\beta
+ d (i_N \omega^\alpha{}_\beta \, \eta_\alpha{}^\beta)\, .
\label{Einham}
\end{equation}
 On a constant
$t=x^0$ surface, the principal term in this Hamiltonian, $-N^\mu
\Omega^\alpha{}_\beta \wedge
\eta_\alpha{}^\beta{}_\mu=N^\mu 2G^\nu{}_\mu \eta_\nu$,
 just reduces to the familiar ADM value $2N^\mu G^0{}_\mu$
(the $i_N\omega$ rotation gauge generator term is
proportional to the torsion).
However the boundary term still needs adjustment \cite{ReTe74}.  We
propose using one of the Hamiltonian boundary terms (\ref{Bomega},
\ref{Btheta}).  Removing the observers ``reference frame
energy-momentum'' $\sim \Lie_N \vartheta \Delta\eta$ via the
identity (\ref{inomega}) gives our covariant quasilocal boundary
expressions for GR:
 \begin{eqnarray} {\cal B}_\omega  &=&
  \Delta \omega^\alpha{}_\beta \wedge i_N  {\buildrel
 \scriptstyle \circ \over \eta}{}_\alpha{}^\beta +
 D_\beta N^\alpha\, \Delta \eta_\alpha{}^\beta \, ,
 \label{Bomega'} \\
{\cal B}_\vartheta  &=&
  \Delta \omega^\alpha{}_\beta \wedge i_N \eta_\alpha{}^\beta +
{\buildrel \scriptstyle \circ \over D}_\beta
{\buildrel \scriptstyle \circ \over N}
 {}^\alpha \, \Delta \eta_\alpha{}^\beta  \, .
\label{Btheta'}
\end{eqnarray}

For the simple alternative choice of a coordinate basis with the
metric and symmetric connection as the variables, the derivation of
the Hamiltonian and its boundary terms is formally the same.  The
control variables are then $\sqrt{-g}g^{\beta\sigma}$ and
$\Gamma^\alpha{}_{\beta\lambda}$ respectively, and the components of
the quasilocal expressions (\ref{Bomega'}, \ref{Btheta'}) take
(with ${\cal B}:= (1/4){\cal B}^{\tau\rho}
 \epsilon_{\tau\rho\mu\nu} dx^\mu \wedge dx^\nu$)
 the form
 \begin{eqnarray}
{\cal B}_{g}^{\tau\rho}
  &=& N^\mu(-g)^{1\over2}\,
g^{\beta\sigma}\,\Delta\Gamma^\alpha{}_{\beta\lambda}\,
 \delta^{\tau\rho\lambda}_{\alpha\sigma\mu}+
  {\buildrel \scriptstyle \circ \over D}{}_\beta
{\buildrel \scriptstyle \circ \over N}{}^\alpha\,\Delta
(\sqrt{-g}g^{\beta\sigma})\,\delta^{\tau\rho}_{\alpha\sigma}\, ,
\label{Bg} \\
{\cal B}_{\Gamma}^{\tau\rho}&=&
 N^\mu (-{\buildrel \scriptstyle \circ \over g})^{1\over2}\,
{\buildrel \scriptstyle \circ \over g}{}^{\beta\sigma}\,
\Delta\Gamma^\alpha{}_{\beta\lambda}\,
\delta^{\tau\rho\lambda}_{\alpha\sigma\mu}+   D_\beta N^\alpha\,
\Delta(\sqrt{-g}g^{\beta\sigma})\,\delta^{\tau\rho}_{\alpha\sigma}\, .
\end{eqnarray}

These new quasilocal boundary expressions are similar to certain
results obtained by earlier investigators.  With a unit displacement,
the first term of ${\cal B}_g$ by itself,
\begin{equation}
U_\mu{}^{\tau\rho}(g,\Delta \Gamma):=
  (-g)^{1\over2}\,
g^{\beta\sigma}\,\Delta\Gamma^\alpha{}_{\beta\lambda}\,
 \delta^{\tau\rho\lambda}_{\alpha\sigma\mu}\, ,
\end{equation}
if we use a vanishing reference connection, reduces to the {\it
Freud superpotential\/} \cite{Fre38}:
 \begin{equation}
U_\mu{}^{\tau\rho}(g,\Gamma)\equiv
(-g)^{-{1\over2}}g_{\mu\nu}\partial_\lambda
\bigl[ -g
(g^{\tau\nu}g^{\rho\lambda}-g^{\rho\nu}g^{\tau\lambda})\bigr]\, .
\end{equation}
Using just this superpotential as the Hamiltonian boundary term is
equivalent to integrating its divergence, the {\it Einstein
pseudotensor\/}, over the spacelike hypersurface.  Our second term,
without the $\Delta$, has the form $2D^{[\rho}N^{\tau]}$, which is the
well known Komar expression \cite{Kom59}.  Several investigators,
working from different perspectives, have arrived at an expression
which is the sum of these two terms:
\begin{equation} \fl\qquad
N^\mu U_\mu{}^{\tau\rho}(g,\Delta\Gamma) +
2\sqrt{-g}\,g^{\mu[\tau}\delta^{\rho]}_\lambda
{\buildrel \scriptstyle \circ \over D}_\mu N^\lambda
\equiv
2\sqrt{-g}\,N^{[\tau}K^{\rho]}+
2\sqrt{-g}\,D^{[\rho}N^{\tau]},
\label{CKS}
\end{equation}
where
$K^\rho:=2g^{\beta[\lambda}\Delta\Gamma^{\rho]}{}_{\beta\lambda}$.
 This was obtained by
Chru\'sciel \cite{Chr85}
and also by Katz \cite{KaOr90} using a background metric while Sorkin
\cite{So88} used only a background connection.  Chru\'sciel's analysis
is especially relevant to us since he also used the Kijowski-Tulczyjew
symplectic-Hamiltonian ideas in much the same spirit as we have done;
his work even includes a second expression corresponding to the
conjugate control mode.  Most of these earlier investigators
recognized the quasilocal possibilities of their results, but they
were primarily concerned with obtaining and discussing the virtues of
this expression for total energy-momentum and angular momentum along
with reconciling the Komar expression with those obtained by other
techniques.  Mention should also be made of an interesting
covariant investigation of Ferraris and Francaviglia \cite{FeFr90}
which uses a global background symmetric connection and has also
arrived at the same expression.

Although the expression (\ref{CKS}) obtained by the aforementioned
investigators
is quite close to our (\ref{Bg}), an important difference
lies in the presence of the $\Delta$ in the ``Komar'' term.  In most
practical quasilocal calculations this amounts to a `difference which
makes no difference'.  However we believe that this $\Delta$ is really
needed for the most general considerations (in particular it is
necessary to get the asymptotically correct behavior for the boundary
term in the Hamiltonian variation).  There are precedents for this
factor beginning with the work of Beig and \'O\ Murchadha
\cite{BeMu87}.

To see this it is necessary to do a space time split of our
expressions.  We `3+1' decompose expression (\ref{Btheta'}) using an
orthonormal frame with one leg $e_\perp$ normal to the hypersurface.
The extrinsic curvature of the spatial hypersurface is related to the
spatial restriction of certain 4-dimensional connection one-form
components:
\begin{equation}
\omega^\perp{}_j=-K_{jl}\vartheta^l\, ,
\end{equation}
while the spatial restriction of the spatial components $\omega^i{}_j$
of the 4-dimensional connection one-form is just the 3-dimensional
connection one-form (for details of the technique see \cite{IsNe80}).
In this way we obtain
 \begin{equation} {\cal
B}_\vartheta  =   \Delta \omega^{ij} \wedge N^\perp \eta_{ij\perp} +
 2 \Delta \omega^{\perp j} \wedge N^k \eta_{\perp jk} +
( {\buildrel \scriptstyle \circ \over D}{}^j
{\buildrel \scriptstyle \circ \over N}{}^\perp
-{\buildrel \scriptstyle \circ \over D}{}^\perp
{\buildrel \scriptstyle \circ \over N}{}^j )
 \, \Delta \eta_{\perp j}\, ,
\label{btheta3+1}
\end{equation}
where we have dropped certain terms, including
$
 {\buildrel \scriptstyle \circ \over D}{}^i
{\buildrel \scriptstyle \circ \over N}{}^j
 \, \Delta \eta_{ij}$,
since they are proportional to
$\vartheta^\perp=N\, dt$ and thus vanish when
restricted to the spatial hypersurface.
Reverting to tensor notation with a coordinate basis, the first term is
\begin{equation}
\Delta\omega^i{}_j\wedge\eta_{i}{}^j{}_{\perp}=\Delta\Gamma^i{}_{jk}
g^{jm}\delta^{lk}_{im}\sqrt{g}dS_l =
 (g^{ij} g^{kl} - g^{il} g^{kj})\,
{\buildrel \scriptstyle \circ \over \nabla}{}_i  \Delta g_{kj}
\sqrt{g}dS_l\, ,
\end{equation}
where we have used
\begin{equation}
\Delta\Gamma^i{}_{kl}={\textstyle{1\over2}}g^{jm}(
{\buildrel \scriptstyle \circ \over \nabla}_k\Delta g_{ml}+
{\buildrel \scriptstyle \circ \over \nabla}_l\Delta g_{mk}-
{\buildrel \scriptstyle \circ \over \nabla}_m\Delta g_{kl})\, .
\end{equation}
Similarly, the second term can be rewritten in terms of the extrinsic
curvature and then in terms of the ADM conjugate momentum:
\begin{eqnarray}\fl
2\Delta\omega^{\perp j}\!\wedge N^k\eta_{\perp jk}&&=
-2\Delta (K^j{}_l\,dx^l)\wedge N^k \eta_{jk\perp}=
-2N^k \Delta K^j{}_m\,\delta^{im}_{jk}\sqrt{g}dS_i \nonumber\\
=-&&2N^k \Delta (K^i{}_k-\delta^i_k
K^j{}_j) \sqrt{g}dS_i=-2N^k \Delta
(-g^{-1/2}\Pi^{ij}g_{jk})\sqrt{g}dS_i\, . \label{supermomentum}
\end{eqnarray}
To appreciate the significance of the remaining term, we consider our
boundary to be at spatial infinity.   Asymptotically, with
${\buildrel \scriptstyle \circ \over N}{}^j$ constant in time and
the usual fall-offs, the final term takes the form
\begin{equation}
 {\buildrel \scriptstyle \circ \over D}{}^j
{\buildrel \scriptstyle \circ \over N}{}^\perp
 \, \Delta\vartheta^k \wedge\eta_{\perp jk}=
g^{jl} N_{,l}\, {\textstyle{1\over2}}\, g^{kn}\Delta g_{ni}\,
\delta^{mi}_{jk} \,\sqrt{g}\, dS_m\, ;
\end{equation}
here we have made the acceptable assumption that $(\Delta
\vartheta^k)_m$ is symmetric and hence determined by $(1/2)\Delta
g_{ij}$.  Taking it all together, at spatial infinity our boundary
term reduces to
 \begin{equation}
\oint d S_l \, \left[\, \sqrt{g}(g^{ij} g^{kl} - g^{il} g^{kj})\,
(N {\buildrel \scriptstyle \circ \over \nabla}{}_i  \Delta g_{kj} +
N_{,i} \Delta g_{kj}) + 2 N^i \Pi^{kl} g_{ik} \,\right],
\end{equation}
where ${\buildrel \scriptstyle \circ \over g}{}_{ij}$ and
 ${\buildrel \scriptstyle
\circ \over \nabla}{}_i$ are flat space quantities.
This is just the Hamiltonian boundary term (generalized to a form
suitable for computation in non-cartesian coordinates) found by Beig
and \'O\ Murchadha \cite{BeMu87} to be necessary to allow for
Poincar\'e displacements of the asymptotic Minkowski structure at
spatial infinity for the asymptotically flat Einstein theory.  The
novel feature which they discovered is the $N_{,i} \Delta g_{jk}$
term, which is, according to their analysis, absolutely necessary to
account for the boosts.  This is where the $\Delta$ in our version of
the ``M{\o}ller-Komar'' term affects the calculated values.
Concerning the value of these boundary terms, the $N$ and $N^k$ terms
give the ADM energy and momentum respectively, the $N^k$ term also
gives the angular momentum.  Note that the $N_{,k}$ term makes no
contribution unless the lapse $N$ is asymptotically unbounded.

More recently Katz, working with Bi{\v c}{\'a}k, Lerer and
Lynden-Bell,
has generalized his previously mentioned work, using Noether
conservation arguments applied to a Lagrangian quadratic in the first
derivatives of the metric along with a global (possibly curved)
background, to obtain a boundary expression for conserved quantities
which can be put in the very succinct form
\begin{equation}
2\{\sqrt{-g}\,N^{[\rho}K^{\tau]}+
\Delta(\sqrt{-g}\,D^{[\rho}N^{\tau]})\}.
\end{equation}
Via manipulations such as those used in conjuction with eq (\ref{CKS})
this is {\it exactly} the same as our expression ${\cal
B}_g^{\tau\rho}$.  We regard this is a significant support for our
particular GR results and thus indirectly for our general ideas
(conversely,
our independent approach also supports the work of these other
investigators).  It seems that our nearly simultaneous efforts are
rather complimentary.  Consequently the successes achieved by these
other workers in applications as diverse as Mach's principle
\cite{LBKaBi95}, conservation laws at null infinity \cite{KaLe97} and
integral constraints for large cosmological perturbations
\cite{KaBiLB97} are further conformations for our work.  This
interesting body of work has only recently come to our attention;
unfortunately, perusal of the connections between our work
and that of Katz et al and the applications suggested thereby cannot
be developed further at this time.

The work just mentioned, however, brings up an alternate
approach to our whole analysis of quasilocal quantities.  Instead of
adjusting the Hamiltonian, we could have begun with an adjustment to
the Hilbert Lagrangian by a boundary term.  Our technique gives
essentially only two different covariant versions for the GR
Lagrangian with boundary term:
\begin{eqnarray}
{\cal L}_{\omega}
  &:=&  \Omega^\alpha{}_\beta \wedge
  \eta_\alpha{}^\beta - d (\Delta \omega^\alpha{}_\beta \wedge
  {\buildrel \scriptstyle \circ \over \eta}{}_\alpha{}^\beta) \, ,
  \\
{\cal L}_{\vartheta}
  &:=&  \Omega^\alpha{}_\beta \wedge
  \eta_\alpha{}^\beta - d (\Delta \omega^\alpha{}_\beta \wedge
  \eta_\alpha{}^\beta) \, .
\end{eqnarray}
Let expressions with a tilde denote the quantity with a trivial
reference configuration, then
\begin{equation}
{\tilde {\cal L}}_{\vartheta}
  :=  \Omega^\alpha{}_\beta \wedge
  \eta_\alpha{}^\beta  - d (\omega^\alpha{}_\beta \wedge
  \eta_\alpha{}^\beta) \,.
\label{Ltilda}
\end{equation}
The boundary term here is just the trace of the extrinsic
curvature---this can be seen by noting that the restriction of the
boundary term to a surface with normal $e_\perp$ is
\begin{equation}
2\omega^\perp{}_a\wedge\eta_\perp{}^a
=-2K_{ab}\vartheta^b\wedge\eta_\perp{}^a=
-2K_{ab}g^{ab}\eta_\perp.
\label{extrinsic}
\end{equation}
Hence we can identify
${\tilde {\cal L}}_{\vartheta}$
as the Lagrangian used by York and Brown \cite{Yo86,BrYo93a};
 its space-time decomposition has the form
\begin{equation}
\fl \qquad S^1 := \int_M {\cal L}_{\mathrm{BY}} =
 {1 \over 2 \chi} \int_M d^4x \sqrt{-g} R
+ {1 \over \chi} \int_{t'}^{t''} d^3x \sqrt{h} K
- {1 \over \chi} \int_{{}^3B} d^3x \sqrt{-\gamma} \Theta\, ,
\end{equation}
where $K$ and $\Theta$ are the traces of the extrinsic curvatures of
the constant $t$ and spatial boundary 3-surfaces which have metrics
$h_{ij}$ and $\gamma_{ij}$, respectively.  Brown and York begin with
this Lagrangian and, without discarding any total derivatives,
construct the Hamiltonian along with its boundary term.  Consequently,
beginning with (\ref{Ltilda}), a straightforward construction of the
`covariant Hamiltonian', without discarding any total derivatives,
leads to their Hamiltonian boundary term expressed in terms of
covariant quantities:
\begin{equation}
{\tilde {\cal B}}_{\mathrm{BY}}(N) =
\omega^\alpha{}_\beta \wedge i_N\eta_\alpha{}^\beta.
\label{tildeBBY}
\end{equation}
By the same type of techniques as used above in connection with
(\ref{btheta3+1},\ref{supermomentum}), it is easily verified that this
4-covariant expression (it's essentially just the Freud
superpotential again) really does yield the densities for the
quasilocal energy and momentum \cite{BrYo93a} which they derived from
the physical Lagrangian $S^1$.

Brown and York took a Hamilton-Jacobi approach;
they chose a reference action and considered $S=S^1-S^0$, so
their renormalized quasilocal quantities are of the form
 \begin{equation}
{\cal B}_{\mathrm{BY}}(N) := {\tilde {\cal B}}_{\mathrm{BY}}(N) -
{\cal B}{}^0_{\mathrm{BY}}(N)\, .
\end{equation}
With a suitable choice for the reference action (it's necessary to use
a reference 2-geometry which is isometric with the boundary 2-surface)
we can obtain a covariant version of their expression:
\begin{equation}
{\cal B}_{\mathrm{BY}}(N) =
  \omega^\alpha{}_\beta \wedge i_N \eta_\alpha{}^\beta-
  {\buildrel \scriptstyle \circ \over \omega}
{}^\alpha{}_\beta \wedge i_N \eta_\alpha{}^\beta
\equiv
  \Delta \omega^\alpha{}_\beta \wedge i_N \eta_\alpha{}^\beta.
\end{equation}
Restricting to the 2-boundary, and using
the previously referred to `3+1' decomposition techniques
such as in (\ref{supermomentum}) along with
 \begin{eqnarray}
\Delta \omega^i{}_j \wedge N^\perp \eta_i{}^j{}_\perp  &=&
2N^\perp\Delta\omega^\vdash{}_A\wedge\eta_\vdash{}^A{}_\perp
=
2N^\perp\Delta( -k_{AB}\vartheta^B)\wedge\eta_\vdash{}^A{}_\perp
\nonumber\\
&=&
-2N^\perp\Delta(k_{AB}\sigma^{AB})\eta_{\vdash\perp}\, ,
\end{eqnarray}
where $e_\vdash$, $\sigma_{AB}$ are, respectively,  the normal and
metric (in an adapted frame) of the 2-boundary,
plus a `2+1' decomposition of (\ref{supermomentum}),  leads
to the quasilocal energy and momentum densities of Brown and York:
 \begin{equation}
\epsilon=k|^{\mathrm{cl}}_0, \qquad
j_i:=-2(\sigma_{ik}(e_\vdash)_l \Pi^{kl}/\sqrt{h})|^{\mathrm{cl}}_0\, ,
\end{equation}
where $k$ is the trace of the extrinsic curvature of the boundary
2-surface embedded in the constant $t$ surface
$\Sigma$.
This covariant version of their expression is closely related to  one
of ours:
 \begin{equation}
 {\cal B}_{\mathrm{BY}}(N)
\equiv
{\cal B}_\vartheta(N)-
i_N{\buildrel \scriptstyle \circ \over \omega}{}^\alpha{}_\beta
 \Delta \eta_\alpha{}^\beta\, .
\label{BBY}
\end{equation}
They exactly agree only when ${\buildrel \scriptstyle
\circ \over \omega}$ vanishes.
Furthermore, the boundary term in the associated
Brown-York Hamiltonian variation,
$d(i_N(\Delta\omega^\alpha{}_\beta\wedge\delta\eta_\alpha{}^\beta)-
i_N{\buildrel \scriptstyle \circ \over \omega}{}^\alpha{}_\beta
 \delta \eta_\alpha{}^\beta) $,
does not have a covariant response unless
${\buildrel \scriptstyle \circ \over \omega}$ vanishes. (In that case
the reference configuration is flat Minkowski space.) From the Komar
form of the difference term, we see that their quasilocal quantities
will have the same values as ours as long as the shift is constant in
time and the lapse is spatially constant.

Although our expressions are quite similar to these well known
expressions of Brown and York, there are some important differences.
In particular
(i) our expressions are manifestly covariant,
(ii) we do not require the boundary to be orthogonal to the spatial
hypersurface (a simplifying restriction they had used which
has recently been relaxed \cite{HaHu96,Lau96,BoMa98}),
(iii) we consider more general reference configurations,
(iv) our displacement vector field $N$ is more
general (as will be explained shortly),
 (v) our expressions have a ``Komar like'' $DN$ term.

%=====================================================================
\subsection{Static spherically symmetric solutions}

As an example and in order to briefly address the issues of the
selection of a reference configuration and displacement vector field,
we consider certain simple solutions of general relativity:  the
static spherically symmetric metrics.  We use the spherical coframe
\begin{equation}
\vartheta^t = \Phi d t, \ \ \vartheta^r = \Phi^{-1} d r, \ \
\vartheta^\theta = r d \theta, \ \ \vartheta^\varphi = r \, \sin \theta
 \,d \varphi,
\end{equation}
where $\Phi=\Phi(r)$.  Specifically we have in mind the
Schwarzschild (anti)-de Sitter and Reissner-Nordstr\"om type metrics:
\begin{equation}
\Phi^2=1-{2M\over r}+\lambda r^2+{Q^2\over r^2}\, .
\end{equation}

In cases like this, where we have an exact analytic form for the metric
which depends on a few parameters, there is an obvious simple choice
for the reference configuration: just allow the parameters to have
their trivial values.  If $\lambda\ne0$ we then have the choice of
a Minkowski or anti-de Sitter reference space.  However it is easy to
see (with computations like those below) that the Minkowski choice
will yield an energy which, for $\lambda\ne0$, diverges as
$r\to\infty$, so we report in detail only the anti-de Sitter choice:
\begin{equation}
{\buildrel \scriptstyle\circ\over \vartheta}{}^t =\Phi_0 d t, \ \
{\buildrel \scriptstyle\circ\over \vartheta}{}^r =\Phi_0^{-1} d r, \ \
{\buildrel \scriptstyle\circ\over \vartheta}{}^\theta =rd\theta, \ \
{\buildrel \scriptstyle\circ\over \vartheta}{}^\varphi =r\, \sin\theta\,
 d \varphi,
\end{equation}
where $\Phi_0^2=1+\lambda r{}^2$.  For the special case of $\lambda=0$
the reference space reduces to Minkowski space.

Because of spherical symmetry, linear and angular momentum vanish, so
we need only calculate the energy.  By symmetry, the timelike
displacement vector field should be orthogonal to the constant $t$
spacelike hypersurfaces.  Thus it should have the form
$N=\alpha e_t$, which still allows for different
``definitions'' of energy.  Our preferred choice is to define energy
using the reference configuration timelike Killing vector
$\partial_{t}=\Phi_0 {\buildrel \scriptstyle \circ \over e}{}_t=\Phi
e_t$, i.e., $\alpha=\Phi$.  Other ``obvious'' candidates are $N =e_t$
and $N= {\buildrel \scriptstyle \circ \over e}{}_t$ which correspond
to the choices of lapse $\alpha=1$ and $\alpha=\Phi_0{}^{-1}\Phi$,
respectively.  We will compare the results of these choices.

Under the conditions we are considering, $\Delta \eta_{tr}$ vanishes,
$i_N\eta_{\alpha}{}^{\beta}=\alpha \eta_{t\alpha}{}^\beta$ and
$i_N {\buildrel \scriptstyle \circ \over \eta}{}_\alpha{}^\beta=
\alpha\Phi^{-1}\Phi_0
{\buildrel \scriptstyle \circ \over \eta}{}_{t\alpha}{}^\beta$.
Consequently
\begin{eqnarray}
{\cal B}_\vartheta(N)
 & =& {\alpha\over 2\chi} \Delta \omega^\alpha{}_\beta \wedge
   \eta{}_{t\alpha}{}^\beta  = {\alpha\over\chi}(
\Delta \omega^{r\theta} \wedge \eta{}_{t r \theta} +
\Delta \omega^{r\varphi} \wedge \eta{}_{t r \varphi}) \nonumber\\
&=& {2\alpha\over\chi} ({\Phi_0-\Phi\over r})\,
\vartheta^\theta\wedge\vartheta^\varphi= {2\alpha\over\chi}
 ({\Phi_0-\Phi})\, r \sin\theta\,  d\theta\wedge d\varphi \, ,
 \end{eqnarray}
which leads to the quasilocal energy
\begin{equation}
E_\vartheta(N)=\alpha\,r\, (\Phi_0-\Phi)\,.
\end{equation}
At the horizon ($\Phi=0$) this reduces to $r\alpha\Phi_0$,
giving
\begin{equation}
E_\vartheta({\buildrel \scriptstyle \circ \over e}_t)=
E_\vartheta(\partial_t)=0, \qquad E_\vartheta(e_t)=r\Phi_0.
\end{equation}
At large distances $E_\vartheta$ approaches
 $(\alpha/\Phi_0)(M-Q^2/2r+\Phi_0^{-2}M^2/2r)$.  Hence, for
$\lambda\ne0$, the asymptotic result is
\begin{equation}
E_\vartheta(\partial_t)=M-Q^2/2r,\qquad
 E_\vartheta(e_t)=E_\vartheta
({\buildrel \scriptstyle \circ \over e}{}_t)=0,
\end{equation}
while for $\lambda=0$
\begin{equation} \!\!\!\!
E_\vartheta(e_t)=M-{Q^2-M^2\over 2r},\qquad
E_\vartheta(\partial_t)=E_\vartheta
({\buildrel \scriptstyle \circ \over e}{}_t)=M-{Q^2+M^2\over 2r}.
\end{equation}

On the other hand, for the other control mode
 \begin{eqnarray} \!\!\!\!
{\cal B}_\omega(N)
 & =& {\alpha\over 2\chi} \Delta \omega^\alpha{}_\beta \wedge
\Phi^{-1}\Phi_0{\buildrel \scriptstyle \circ \over \eta}
{}_{t\alpha}{}^\beta  = {\alpha\Phi_0\over\chi\Phi}(
\Delta \omega^{r\theta} \wedge
 {\buildrel \scriptstyle \circ \over \eta}{}_{t r \theta} +
\Delta \omega^{r\varphi} \wedge
{\buildrel \scriptstyle \circ \over \eta}{}_{t r \varphi}) \nonumber\\
&=& {2\alpha\Phi_0\over\chi\Phi}({\Phi_0-\Phi\over r})\,
{\buildrel \scriptstyle \circ \over \vartheta}{}^\theta\wedge
{\buildrel \scriptstyle \circ \over \vartheta}{}^\varphi
= {2\alpha\Phi_0\over\chi\Phi}
({\Phi_0-\Phi})\, r \sin\theta \,
d\theta\wedge d\varphi \, ,
 \end{eqnarray}
leads to the quasilocal energy
\begin{equation}
E_\omega(N)=
\alpha\,r\, \Phi^{-1}\Phi_0(\Phi_0-\Phi)\,.
\end{equation}
At the horizon this is $r\alpha(\Phi_0^2/\Phi)$, giving
\begin{equation}
E_\omega(e_t)=\infty, \quad
E_\omega(\partial_t)=r\Phi^2_0=2(M-Q^2/2r),\quad
E_\omega({\buildrel \scriptstyle \circ \over e}{}_t)=r\Phi_0,
\end{equation}
while for very large $r$ it approaches
 $(\alpha/\Phi)(M-Q^2/2r+\Phi_0^{-2}M^2/2r)$, giving, for
$\lambda\ne0$,
 \begin{equation}
E_\omega(\partial_t)=M-Q^2/2r, \qquad
E_\omega({\buildrel \scriptstyle \circ \over e}{}_t)=
E_\omega(e_t)=0,
\end{equation}
and, for $\lambda=0$,
\begin{equation}
E_\omega({\buildrel \scriptstyle \circ \over e}{}_t)=
E_\omega(\partial_t)=M-{Q^2 - M^2\over 2r},\qquad
E_\omega(e_t)=M-{Q^2 -3M^2\over 2r}.
\end{equation}
Because of the vanishing values at large distances for $\lambda\ne0$,
$e_t$ and ${\buildrel \scriptstyle \circ \over e}{}_t$ seem to be
unsuitable choices.  Then the vanishing of $E_\vartheta(\partial_t)$
at the horizon suggests that analytic matching may not be so
physical.

A reasonable alternate way to determine the values of the reference
quantities is by embedding a neighborhood of the boundary into a space
which has the desired reference geometry and then pulling back all the
quantities to the dynamic spacetime.  In the present case
we can assume that the reference geometry coframe has the same
spherical form: \begin{equation}
{\buildrel \scriptstyle\circ\over \vartheta}{}^t =\Phi_0 d t', \ \
{\buildrel \scriptstyle\circ\over \vartheta}{}^r =\Phi_0^{-1} d r', \ \
{\buildrel \scriptstyle\circ\over \vartheta}{}^\theta =r'd\theta', \ \
{\buildrel \scriptstyle\circ\over \vartheta}{}^\varphi =r'\,
 \sin\theta'\, d \varphi',
\end{equation}
where, in particular, $\Phi_0^2=1+\lambda r'{}^2$.  We first identify
corresponding foliations by spacelike hypersurfaces.  A good criteria
for this identification is to relate points that have the same trace
for the extrinsic curvature.  In this particular case this criteria is
satisfied by the obvious choice:  the one form $dt'$ corresponds to
$\beta dt$ for some function $\beta$.  Within each spacelike
hypersurface a neighborhood of the (assumed to be spherical) boundary
$S=\partial \Sigma$ is diffeomorphic to a suitable image in the
reference geometry.  A reasonable choice for the embedding is to
require that the corresponding 2 spheres have the same {\em intrinsic}
geometry.  The uniqueness of such an identification has been discussed
in \cite{BrYo93a}.  (One alternative would be to try to match the {\em
extrinsic} geometries.) \ \ In the present case it simply means that
$r=r'$, $\theta=\theta'$, $\varphi=\varphi'$.  To completely fix the
embedding we need to specify the corresponding timelike unit.  An
obvious simple choice is to take $t=t'$, i.e., $\beta=1$.  That just
gives the analytic matching case already considered.  On the other
hand, giving due consideration to the operational measurement
procedure, it seems more physically reasonable as well as more
geometric to match corresponding units of {\em proper} time rather
than coordinate time.  Thus we can choose to fix our relative time
coordinates by identifying $\vartheta^t$ and
${\buildrel \scriptstyle \circ \over \vartheta}{}^t$ on
$\partial\Sigma$,  consequently
$\beta=\Phi\Phi_0^{-1}$.  It should be noted, however, that this
choice {\em is not} integrable.  This type of geometric matching is
{\em instantaneous}.  At a later instant of time the procedure will
determine a new reference configuration.  It cannot be expected that
these instantaneous reference configurations will mesh to form a
single reference geometry.

Proceeding with our quasilocal calculations as before, again, because
of spherical symmetry we need only calculate the energy and
different choices for the displacement vector field give
different ``definitions'' of the
energy. Our preferred choice is to  use the reference
configuration timelike Killing vector $\partial_{t'}$ which
corresponds in this case to the choice of lapse $\alpha=\Phi_0$.
Other obvious candidates are $N = \partial_t$
and $N=e_t={\buildrel \scriptstyle \circ \over e}{}_t$
which correspond respectively to the choices of lapse $\alpha=\Phi$ and
$\alpha=1$.  We will now compare the results of these choices.  Under
the conditions we are considering, $\Delta \eta_{tr}$ vanishes and
$i_N\eta_{\alpha}{}^{\beta}=\alpha \eta_{t\alpha}{}^\beta=
\alpha{\buildrel \scriptstyle \circ \over \eta}{}_{t\alpha}{}^\beta=
i_N{\buildrel \scriptstyle \circ \over \eta}{}_\alpha{}^\beta$
so
\begin{eqnarray}
{\cal B}_{\omega}={\cal B}_{\vartheta}
 & =& {\alpha \over 2 \chi} \Delta \omega^\alpha{}_\beta \wedge
   \eta{}_{t\alpha}{}^\beta
  = {\alpha \over  \chi} (
\Delta \omega^{r\theta} \wedge \eta{}_{t r \theta} +
\Delta \omega^{r\varphi} \wedge \eta{}_{t r \varphi}) \nonumber\\
&=& {2\alpha\over\chi} ({\Phi_0-\Phi\over r})\,
\vartheta^\theta\wedge\vartheta^\varphi= {2\alpha\over\chi}
 ({\Phi_0-\Phi})\, r \sin\theta \, d\theta\wedge d\varphi \, .
 \end{eqnarray}
Consequently the quasilocal energy is
\begin{equation}
E(N)=E_\vartheta(N)=E_\omega(N)=\alpha\,r\, (\Phi_0-\Phi)\,.
\end{equation}
For very large $r$ this approaches
$\alpha(\Phi_0^{-1}(M-Q^2/2r)+\Phi_0^{-3}M^2/2r)$
while at the horizon ($\Phi=0$) it reduces to $\alpha r \Phi_0$.
Hence, for asymptotically Minkowski spaces ($\lambda=0$, $\Phi_0=1$)
our expressions yield
\begin{equation}
E(\partial_{t'})=E(e_t)=M+{M^2-Q^2\over 2r}\, ,\qquad
E(\partial_t)=M-{M^2+Q^2\over 2r}\, ,
\end{equation}
for very large $r$ and
\begin{equation}
E(\partial_{t'})=E(e_t)=r, \qquad E(\partial_t)=0,
\end{equation}
at the horizon.
For asymptotically anti-de Sitter spaces our expressions yield
\begin{equation}
E(\partial_{t'})=E(\partial_t)=M-{Q^2\over 2r}\, ,\qquad
E(e_t)=M\Phi_0^{-1}\to0,
\end{equation}
for very large $r$ and
\begin{equation}
E(\partial_{t'})=r\Phi_0^2, \qquad E(\partial_t)=0,
\qquad E(e_t)=r\Phi_0,
\end{equation}
at the horizon.

We have proposed a reference configuration Killing field as a good
choice for the displacement vector field.  The above calculations show
that for geometric matching this choice gives reasonable values in
practice.  Note that the choice $N=\partial_t$ gives a vanishing value
for the quasilocal energy within the horizon.  For asymptotically flat
space our spherically symmetric quasilocal energy values are the same
as those found by Brown and York from their Hamilton-Jacobi approach.
More generally, our proposal for the evolution vector field relaxes
their choice of $\alpha=1$.  Consequently we get the total energy $M$
for asymptotically anti de-Sitter solutions whereas their choice of
$\alpha=1$ leads to $E=M(1+\lambda r^2)^{-1}$ which vanishes
asymptotically.  For the Hamilton-Jacobi approach to anti de-Sitter
space see \cite{BrCrMa94}

A direct consequence of our favored choice of the reference
configuration timelike Killing vector is that, just as for Brown and
York, the quasilocal energy for the Schwarzschild solution is $2M$ at
the horizon and $M$ at infinity.  Thus between the horizon and spatial
infinity the quasilocal energy is {\em negative}, contrary to a
property favored by several investigators \cite{ChYa88, DoMa91,
Be92b}.  However, as G.~Hayward \cite{HayG95} has observed, the
quasilocal energy {\em cannot} be positive in general---simply because
closed spaces must have zero total energy so they must have negative
regions to balance the positive regions.

The result of  Landau-Lifshitz, Tolman
\cite{Vir90a} and S. Hayward \cite{HayS94} for the
Reissner-Nordstr\"om metric is $E = M - Q^2 / 2 r$. They find
that the gravitational field has a remarkable difference from the
electromagnetic field.  The energy of the electromagnetic field is
shared by the interior as well as the exterior of the horizon of the
system, but the gravitational field energy is confined to its interior
only.  However, in our result the distribution of the gravitational
and electromagnetic energy contributions is similar.  Both are shared
by the interior and exterior.  One again this is a consequence of our
choice for $N$.  For the $Q=0$ case, some, e.g.  \cite{ChYa88}, have
argued that the quasilocal energy for spherically symmetric solutions
should simply be $M$ independent of $r$.  In our formulation that can
be achieved simply by choosing the lapse to have a suitable value.
(The special orthonormal frame Hamiltonian \cite{Ne91b} also naturally
suggests a choice which ``localizes'' all of the mass within the
horizon.) More generally we can obtain the aforementioned value
$E = M - Q^2/2r$ simply by choosing a (rather strange) value for the
lapse such as $\alpha=(\Phi_0+\Phi)/2$.  By a judicious choice of the
reference configuration as well as the lapse our expressions can yield
results matching most of the values (e.g., those in \cite{PeNa96})
found by other investigators for spherically symmetric systems.

%%%%%%%%%%%%%%%%%%%%%%%%%%%%%%%%%%%%%%%%%%%%%%%%%%%%%%%%%%%%%%%%%%%%%%
% Discussion                                                         %
%%%%%%%%%%%%%%%2%%%%%%%%%%%%%%%%%%%%%%%%%%%%%%%%%%%%%%%%%%%%%%%%%%%%%%%
\section{Discussion}

The quest for gravitational energy-momentum has led to the quasilocal
idea:  {\em quasilocal quantities}, including energy-momentum and
angular momentum, are associated with a closed 2-surface.  In this
work we have taken a Hamiltonian approach, defining the quasilocal
quantities in terms of the value of the Hamiltonian for a finite
region.  This value is completely determined by the Hamiltonian
boundary term.

We have considered a rather general class of geometric gravity
theories, namely those having an independent metric and connection
(Metric-Affine gravity).  The field equations were presumed to follow
from a Lagrangian which depends on the metric, torsion and curvature.
(However, there is no real barrier to extending the procedure to
include
derivatives of these fields.)\ \ In our approach, special restrictions
including metric compatibility, a symmetric connection or teleparallel
geometry are easily introduced to lead to the familiar theories of
interest to most investigators.

For this general class of geometric gravity theories, by using a
covariant canonical Hamiltonian formalism, we presented both a general
procedure, based on the boundary term in the variation of the
Hamiltonian and its symplectic structure, and a specific proposal:
that this term should have a well defined {\em covariant} symplectic
structure.  We found that, for each dynamic field, there is a choice
between two covariant boundary terms (essentially Dirichlet or Neumann
boundary conditions).  We then applied our formalism to black hole
thermodynamics obtaining thereby an expression for the entropy in
these general theories.  Then we restricted our expressions to general
relativity and applied them to well known spherically symmetric
solutions, comparing the results with those found by some other
authors.  (Remarkably one of our expressions turns out to be
equivalent to an expression recently developed and applied by Katz
and his coworkers.)\ \  An important omission, in the context of
Einstein's theory,
is the relationship of our formalism to spinor formulations.  This
topic will be considered in a follow up work \cite{ChNe98}.  Here we
wish to further discuss several issues that have arisen.

From a physical (``operational'') point of view, suppose some
observers attempt to measure the quasilocal quantities.  What kind of
procedure could they use?  What will the result depend on?  Ideally it
should be purely geometric --- depending only on the dynamical
variables at the boundary, independent of any reference frame or
choice of coordinates.  Careful analysis, however, has led us to the
recognition that specifying a quasilocal gravitational energy-momentum
requires many choices, in particular
\begin{itemize}
\item[(i)] the theory: e.g., GR, Brans-Dicke, Einstein-Cartan, PGT, MA
(Metric-Affine \cite{HeMcMiNe93});
\item[(ii)] the representation or dynamic variables: e.g., the metric,
orthonormal frame, connection, spinors;
\item[(iii)] the control mode or boundary conditions: e.g., covariant,
Dirichlet/Neumann for each dynamic variable;
\item[(iv)] the reference configuration: e.g., Minkowski;
\item[(v)] a displacement vector field: we propose using a
Killing field of the reference geometry.
\end{itemize}
In this investigation we have determined how the
quasilocal expressions should depend on these choices.

Given a specific theory and representation our formalism determines
the covariant quasilocal expressions uniquely, however we do not yet
get a unique value for the quasilocal quantities in a finite region.
This is because there are two features still to be determined.  One is
the displacement vector $N$, and the other is the reference
configuration.

The most ambiguous part of our program is the reference configuration.
(We prefer the more general term ``reference configuration'' as
opposed to ``reference geometry'' simply because the same type of
analysis applies to other dynamic fields, e.g., electromagnetism.)\ \
It can not be avoided in our expressions.  Operationally it can be
related to the calibration or ``renormalization'' of our measurement.
If the dynamical variables take on the reference values our
expressions will give zero values for all quasilocal quantities.

How to choose an appropriate reference configuration?  We are not
really satisfied with our understanding of this topic but will share
our current thoughts here.  Since our boundary could be anywhere we
could simple use a globally fixed background geometry.  For an
asymptotically flat or constant curvature spacetime, we can choose the
simplest global spacetime (let the source vanish in the original
spacetime) which has the same asymptotic behavior as the physical
spacetime.  This is not so ambiguous asymptotically; it allows one to
determine the total conserved quantities.  But a global background is
hardly necessary, moreover it is operationally impractical.  How then
should one determine the reference configuration for a finite region?

We can imagine the dynamic fields in the neighborhood of the boundary
as deviating from some ``ground state'' configuration.  These
``reference values'' can be viewed as defining (at least a portion of)
a conceptually distinct geometric reference space.  A slight shift of
view then leads to a simple method (which we have elaborated upon) for
constructing the reference configuration for the geometric quantities
of interest here:  use an embedding (this depends on some matching
criteria such as proper time and intrinsic 2-surface isometry) of a
certain neighborhood of the topologically compact orientable boundary
spatial 2-surface (usually with spherical topology) into a space
having the desired reference geometry (ordinarily a Minkowski
geometry, but alternatives include Schwarzschild, (anti) de Sitter, or
a homogeneous cosmology).  Pull back the geometry to determine the
reference configuration.  (At the same time we can also pull back a
displacement vector.)\ \ The important question is whether we can find
a unique embedding?  (And, of course, what does it mean physically?)\
\ Brown and York have proposed embedding the 2-surface isometrically
\cite{BrYo93a}.  They have referred to some uniqueness theorems for
the embedding of topologically spherical positive curvature 2-metrics.
Such an embedding seems like a good idea.  However, we do not yet
adequately understand what this means physically or whether some other
idea would be equally reasonable.

Another way to deal with the reference configuration is to replace it
with some additional covariant field satisfying some propagation
equation.  To our knowledge this approach has so far only been used in
the context of spinor expressions \cite{Sz94, La95}.  In this case the
propagation equation, in effect, implicitly determines the reference
configuration.  For example, the Dougan and Mason spinor propagation
equation \cite{DoMa91} determines a spinor field on a closed
2-surface.  There is a good prospect for relating this approach and the
geometric embedding approach.  For example a spinor field satisfying a
suitable propagation equation could be used to determine an
orthonormal frame which, in turn, could be used to determine a
reference configuration for the geometrical quantities.

Turning to the displacement vector $N$ (which corresponds to the {\it
lapse function} and the {\it shift vector} in the ADM analysis
\cite{ArDeMi62}).  How should we choose the appropriate displacement
vector in order to get the ``proper'' quasilocal quantities?  For the
total conserved quantities, the choice is unique, namely an asymptotic
Killing vector.  But for the quasilocal quantities in a finite region,
the corresponding Killing vector may not exist.  Obviously energy is
related to time translation, momentum to space translation and angular
momentum to rotation but which precise choice of timelike vector field
gives the energy?  For example the Brown and York \cite{BrYo93a}
choice for energy, lapse $=1$ and shift $ = 0$, seems natural.  Yet
with it, for asymptotically anti-de Sitter solutions, the quasilocal
energy will not converge to the total energy.  The choice of $N$ is
probably best tied to the choice of reference configuration.  We have
proposed using a Killing field of the reference configuration to
define quasilocal energy, momentum and angular momentum.  As in
\cite{BrYo93a} alternate choices for the displacement vector field
could be used to distinguish a quasilocal quantity from a conserved
charge.  For example, each Killing field of the dynamic spacetime
would give a conserved charge.

The quasilocal expressions that we found are well behaved and are
related to expressions found by other investigators.  They satisfy the
usual criteria:  their values have good correspondence limits to flat
space and spherically symmetric values, with good weak field and
asymptotic limits for both asymptotically flat and constant curvature spaces,
and in both the Bondi and ADM limit.  (However a complete analysis of the
asymptotics which will give well defined quasilocal quantities and the
necessary conditions for the Hamiltonian to be a differential generator in the
sense of \cite{BrHe86} remains to be completed.)\ \ Although our
expressions mesh with a positive
total energy proof for asymptotically flat solutions to Einstein's theory,
our expressions are not locally positive in general.  However, we
regard this as a virtue rather than a drawback for they thus allow for
the correct (vanishing) total for closed spaces.  In any case, such
criteria are known to be insufficient.

We have emphasized an additional principle of the Hamiltonian
formalism concerning the vanishing of the boundary term in the
variation of the Hamiltonian for suitable boundary conditions.
Specifically we have required that the boundary term in the
Hamiltonian variation have a well defined {\em covariant symplectic}
structure.

We stated that we have determined the expressions for quasilocal
quantities uniquely.  Here the meaning of the word {\it uniquely} is
based on the requirement of the covariant symplectic structure.  We
could release this rather strong requirement to a weaker version which
only requires a well defined symplectic structure.  For example, for a
vector field $W^\mu$, we could use Dirichlet conditions on part of the
spacetime components say $W^0, W^1$ and Neumann conditions on the
remaining components.  Then many more quasilocal expressions are
possible.  This is just the sort of thing that happens with the
$i_N$ factor in the boundary term in the Hamiltonian variations
(\ref{C(N)1},\ref{C(N)'}).

From the physical point of view, certain noncovariant
choices might be distinguished and may sometimes be more practical.
One approach is to consider the symplectic structure, taking into
account the initial value constraints or the boundary conditions
needed for a good initial value problem.  (Boundary control or fixed
on the boundary, by the way, does not mean having a constant value,
but rather means that the value of the variable is a preprogrammed
function of time.) In particular, for the Einstein theory, Jezierski
and Kijowski \cite{JeKi90} have investigated decomposing the variables
into the true unconstrained degrees of freedom.  They considered the
form of the initial value constraints and found that for certain
control modes the boundary value problem was not well posed.  From a
naive view point, the symplectic structure reflects the
control-response relation of the physical system.  So for some
specific theory, the symplectic structure may need some modifications,
because the constraints could forbid some control mode at the
boundary.  In principle, we could handle such constraints by imposing
them with Lagrange multiplier terms in the variational principle.
This may modify the symplectic structure and the expressions for
quasilocal quantities, effectively automatically including the
constraints.  Further investigation is certainly needed on this
topic.

Others boundary control choices might be more closely related to what
an observer would directly measure.  For example, in an
electromagnetic system, we can naturally fix the electric potential
and the tangential components of the magnetic field instead of the
whole 4-vector potential at the boundary, or we can fix the normal
component of the electric field and the tangential components of the
vector potential (essentially the normal component of the magnetic
field).  The physical meaning as well as the procedure for
implementing these choices is well understood.  But the boundary term
in the corresponding Hamiltonian variation for such control modes
depends on the fields in a noncovariant way.  However, these
non-covariant expressions are related to a covariant expression of our
type by a simple Legendre transformation on the boundary.  Having
measured the values for one of these situations, we could calculate
the value of our covariant quasilocal expressions even if they could
not be conveniently measured directly.

We have emphasized the importance of covariance for good physical
quasilocal quantities. Yet here exists a noncovariant term,
$i_N \omega^\alpha_{\ \beta}\Delta \rho_\alpha^{\ \beta}$,
in our Hamiltonian boundary expressions.  This term merits deeper
consideration.  We noted that it was through this term that our
expressions mainly differed from those found by others.  Consequently,
further investigation is desirable to ascertain its importance and how
essential it is that it have exactly the form that we found.  We have
decomposed this term by using the identity (\ref{inomega}).  The first
two terms on the rhs of this identity are covariant and represent a
real physical effect.  The last term, which reflects a noncovariant
property of our expressions, is an unphysical term presenting a
dynamical reference frame effect.  This situation is like the
centrifugal force in mechanics.  The complete Hamiltonian generates
the (arbitrary) dynamical evolution of the frame (effectively, the
observer) as well as the physical variables.  Through the relation
(\ref{inomega}), the complete Hamiltonian implicitly includes a time
derivative term on the boundary.  Within the context of Einstein's
theory, boundary time derivative terms have been noted some time ago
by Kijowski \cite{Ki83} and more recently by G.  Hayward
\cite{HayG95}.

A major virtue of our formalism is that it provides for system and
order.  Although it allows for many ``energies'', yet each has (in
principle) a clear physical meaning.  The formalism systematically
associates a quasilocal energy-momentum with a specific Hamiltonian
boundary term (determined by the choice of theory,
representation---dynamic variables, control mode, reference
configuration, matching criteria, and displacement vector field).  The
gravitational quasilocal energy-momentum is thus, like the energies in
thermodynamics, connected with a definite physical situation with
definite conditions on the boundary.  But much more investigation is
needed to understand the significance of various boundary conditions.
Hence our formalism shifts the attention to questions like ``What is
the real physical significance of controlling the metric (or the
connection) on the boundary of a finite region?  Which boundary
control (if any) is the analog of thermally insulating a system?
Which corresponds to a thermal bath?''  Some insight is afforded by
classical electrodynamics where the answers to such questions are
understood \cite{Ki83, JeKi90}.

Of course one way to further investigate the various quasilocal
expressions is to do more direct calculations for exact solutions,
e.g., \cite{Mar94}.  However, a deeper theoretical investigation
should be more revealing.  Our formulation provides a good starting
point for such an investigation.  Note that {\em all\/} of the
expressions presented here correspond to the work done in some (ideal)
physical process.  The situation is similar to thermodynamics with its
different energies (enthalpy, Gibbs, Helmholtz, etc.)\ \ An even
better analogy is the electrostatic work required while controlling
the potential on the boundary of a region vs.\ that required while
controlling the charge density \cite{KiTu79}.

However, we have gone far enough for the present.  Our intention here
was primarily to set up a framework, to indicate something of its
scope and the efficacy of the principles.  In follow up works we plan
to extend these investigations in several directions including the
application to and relationship with spinor expressions for quasilocal
quantities \cite{ChNe98}.

%%%%%%%%%%%%%%%%%%%%%%%%%%%%%%%%%%
% Acknowledgment
%%%%%%%%%%%%%%%%%%%%%%%%%%%%%%%%%%
\section*{Acknowledgments}

We would like to express our appreciation for input and several
good suggestions from the referee, J. Katz and M. Godina
as well as to R.S. Tung for his
advice and assistance.
This work was supported by the National Science Council of the R.O.C.
under grants No. NSC87-2112-M-008-007,
NSC86-2112-M-008-009, NSC85-2112-M008-003,
NSC84-2112-M-008-004 and NSC83-0208-M-008-014.

%%%%%%%%%%%%%%%%%%%%%%%%%%%%%%%%%%%%%%%%%%%%%%%%%%%%%%%%%%%%%%%%%%%%%%
% References --- in cited order                                      %
%%%%%%%%%%%%%%%%%%%%%%%%%%%%%%%%%%%%%%%%%%%%%%%%%%%%%%%%%%%%%%%%%%%%%%

\end{document}